\documentclass[english,aps,superscriptaddress,prl,twocolumn]{revtex4-2}
\usepackage{amsmath}
\usepackage{amssymb}

\usepackage{newtxmath}
\usepackage[LGR,T1]{fontenc}
\usepackage[latin9]{inputenc}
\usepackage{color}
\usepackage{pifont}
\usepackage{bbding}
\usepackage{graphicx}

\makeatletter

\providecommand{\tabularnewline}{\\}

\renewcommand{\figurename}{FIG.}
\usepackage[colorlinks=true,linkcolor=blue,citecolor=blue,urlcolor=blue]{hyperref}

\newcommand*\LyXThinSpace{\,\hspace{0pt}}
\DeclareRobustCommand{\greektext}{%
  \fontencoding{LGR}\selectfont\def\encodingdefault{LGR}}
\DeclareRobustCommand{\textgreek}[1]{\leavevmode{\greektext #1}}

\makeatother

\usepackage{babel}
\begin{document}
\renewcommand{\figurename}{FIG.}
\title{Altermagnetism Induced Bogoliubov Fermi Surfaces Form Topological
Superconductivity}
\author{Bo Fu}
\email{fubo@gbu.edu.cn}

\affiliation{School of Sciences, Great Bay University, Dongguan, China}
\author{Chang-An Li}
\email{changan.li@uni-wuerzburg.de}

\affiliation{Institute for Theoretical Physics and Astrophysics, University of
Würzburg, 97074 Würzburg, Germany}
\affiliation{Würzburg-Dresden Cluster of Excellence ctd.qmat, Germany}
\author{Björn Trauzettel}
\affiliation{Institute for Theoretical Physics and Astrophysics, University of
Würzburg, 97074 Würzburg, Germany}
\affiliation{Würzburg-Dresden Cluster of Excellence ctd.qmat, Germany}
\date{\today}
\begin{abstract}
We propose a novel type of topological superconductivity based on
Bogoliubov Fermi surfaces (BFSs) in an altermagnetic topological insulator
proximitized by an $s$-wave superconductor. The 3D altermagnetic
topological insulator is characterized by zero-energy surface states
in bulk nodal-ring phases and anisotropically shifted surface Dirac
cones in topological insulating phases. The altermagnetic order in
combination with superconductivity gives rise to highly anisotropic
superconducting gaps with crystal-facet-dependent BFSs at the physical
boundaries. These particular BFSs provide distinct platforms to realize
topological superconductivity. We propose a quasi-1D nanowire in which
the anisotropic BFSs experience topological phase transitions due
to quantum confinement leading to Majorana zero modes (MZMs) at its
ends. We further consider vortex phase transitions in the superconducting
altermagnetic topological insulators. Remarkably, we find that the
altermagnetic order allows us to transit between two distinct type
of MZMs, one type is located at the vortex line, while the other type
is located at the physical boundaries. Our work paves a new avenue
utilizing altermagnetism-induced BFSs to engineer topological superconductivity
through crystal anisotropy and quantum confinement.
\end{abstract}
\maketitle
\textit{\textcolor{blue}{Introduction.---}}The recent discovery of
altermagnets (AMs) has introduced a new type of magnetic phase characterized
by momentum dependent spin splitting but with zero net magnetization
\citep{Smejkal22prxc}. It bridges the gap between conventional ferromagnetic
and antiferromagnetic phases offering a fertile playground for novel
physics and promising applications \citep{Naka19NC,Smejkal20SACrystal,ShaoDF21NC,fukaya2025superconducting,Smejkal22prxc,Libor22prx2,BaiL24AFM,GuMQ25PRL,Song25NRM,Hayami19JPSJ}.
Extensive research in recent years has been dedicated to their fundamental
classification \citep{liu2022spin,xiao2024spin,yang2024symmetry}
and magnetic/electronic properties \citep{steward2023dynamic,fernandes2024topological,yu2024spin,fang2024quantum,autieri2025staggered}.
Crucially, the zero net magnetization in AMs significantly eliminates
detrimental stray fields that typically suppress superconducting order
while the strong spin splitting is essential for manipulating electronic
states \citep{blamire2014interface,parshukov2025exotic}. Since no
intrinsic superconducting AM has been found so far, it is important
to explore if AMs proximitized to conventional $s$-wave superconductors
result in topological phases and phenomena \citep{Beenakker23prb,alam2025proximity,zhu2025altermagnetic,wei2024gapless,giil2024superconductor,lu2025engineering,nagae2025flat,heinsdorf2025proximitizing}.
Related investigations of AM-superconductor (SC) hybrids have made
substantial progress in relation to Andreev reflection \citep{SunC23prb,beenakker2023phase,Papaj23PRB},
Josephson effects \citep{Ouassou23PRL,lu2024varphi,HPSun25PRB,maeda2024theory,ZhangSB24NC},
and novel topological phases \citep{DiZhu23PRB,zou2024topological,ezawa2024topological,zhang2024topological,ma2024altermagnetic,gonzalez2025model,gonzalez2025spin,li2025floating,chen2025quantum,Antoneko25prl,chen2025probing,li2025altermagnetism,heinsdorf2025altermagnetic,wan2025altermagnetism}.
In particular, the interplay between altermagnetism and topology presents
possibilities to realize topological superconductors and MZMs without
external fields \citep{Ghorashi24PRL,li2024realizing,li2024creation}.
However, most research efforts have been put into the altermagnetic
order in the bulk states, the influence of altermagnetism on topological
surfaces states remains a largely unexplored avenue. Indeed, a critical
open question is whether the particular anisotropic features of AMs
can be utilized to engineer topological surface states yielding topological
superconductivity.

\begin{figure}
\includegraphics[width=8cm]{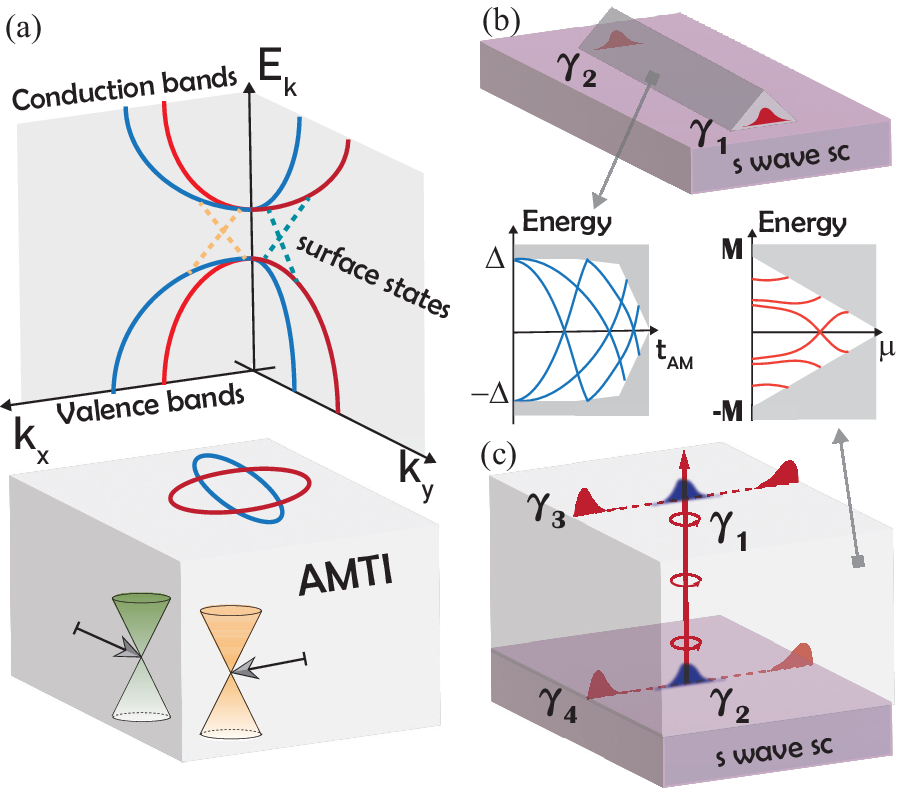}

\caption{(a) Schematic of the anisotropic surface states in an AMTI. (b) A
quasi-1D triangular prism nanowire of AMTI/SC heterostructure hosting
a pair of MZMs at the two ends. The spectrum illustrates the evolution
of BFSs to energy levels (blue lines) driven by AM order strength
$t_{AM}$ inside the superconducting gap $\Delta$. (c) AMTI/SC hybrid
structure with a vortex line, hosting MZMs at vortex core ($\gamma_{1,2}$)
or side surfaces ($\gamma_{3,4}$), depending on the value of $t_{AM}$.
The energy spectrum depicts the evolution of BFSs to discrete surface
energy levels (red lines) as a function of chemical potential $\mu$
within the bulk band gap. \label{Schematic_diagram}}
\end{figure}

In this work, we directly address this question by demonstrating that
it is possible to realize topological superconductivity via topological
phase transitions of surface states in altermagnetic topological insulators
(AMTIs) in proximity to $s$-wave superconductors. As illustrated
in Fig. \ref{Schematic_diagram}(a), AMTIs are TIs endowed with $d$-wave
altermagnetic order, which induces a strong momentum-dependent spin
polarization in their bulk bands. Consequently, new topological phases
such as bulk nodal-ring semimetals appear in AMTIs. The anisotropy
of AMs is further imprinted on the topological surface states, where
the Dirac cones exhibit a pronounced and facet-dependent momentum
shift. When proximitized with an $s$-wave superconductor, the spin-splitting
suppresses conventional singlet pairing between time-reversed states.
This leads to the formation of crystal-facet-dependent BFSs at the
side surfaces, which constitutes the basis for realizing topological
superconductivity and MZMs in our proposal. In a prism nanowire geometry,
quantum confinement further discretizes the surface BFSs into a series
of energy levels within the superconducting gap {[}see Fig. \ref{Schematic_diagram}(b){]}.
The altermagnetic order thus acts as a knob to drive topological phase
transitions of these surface states, ultimately generating MZMs at
the wire ends in nontrivial phases. We further demonstrate that altermagnetic
order provides a unique mechanism to exert effective control over
vortex-induced MZMs {[}see Fig. \ref{Schematic_diagram}(c){]}. Notably,
the altermagnetic order drives not only vortex phase transitions but
also topological phase transitions of BFSs at side surfaces, which
is beyond the vortex-line physics in conventional superconducting
TIs. This transition forces MZMs to migrate from the vortex core to
the physical boundaries. Our results provide highly tunable platforms
for realizing and controlling topological superconductivity and MZMs
by exploiting the profound impact of altermagnetism on topological
surface states.

\textit{\textcolor{blue}{Model and general analysis.---}}We start
with a 3D lattice model describing the AMTI proximitized to an $s$-wave
superconductor. The corresponding Bogoliubov--de Gennes (BdG) Hamiltonian
constructed in the Nambu basis $\Psi_{\mathbf{k}}^{\dagger}=(C_{\mathbf{k}}^{\dagger},C_{-\mathbf{k}}^{\mathrm{T}}i\sigma_{y})$
takes the form:
\begin{equation}
\mathcal{H}_{\mathrm{BdG}}(\mathbf{k})=\left(\begin{array}{cc}
h(\mathbf{k})-\mu & \Delta\\
\Delta^{*} & -\sigma_{y}h^{*}(-\mathbf{k})\sigma_{y}+\mu
\end{array}\right),
\end{equation}
where $\Delta$ is the superconducting pairing potential, $\mu$ the
chemical potential, and ${\bf k}=(k_{x},k_{y})$ the Bloch wavevector.
The normal-state Hamiltonian $h(\mathbf{k})$ for the 3D AMTI is given
by \citep{ma2024altermagnetic,gonzalez2025model}
\begin{align}
h(\mathbf{k})= & M(\mathbf{k})\rho_{z}\sigma_{0}+\lambda(\sin k_{x}\rho_{x}\sigma_{x}+\sin k_{y}\rho_{x}\sigma_{y})\nonumber \\
 & +\lambda\sin k_{z}\rho_{x}\sigma_{z}+t_{AM}(\cos k_{x}-\cos k_{y})\rho_{z}\sigma_{z}.\label{eq:Hamiltonian}
\end{align}
The Pauli matrices $\sigma_{\alpha=x,y,z}$ and $\rho_{\alpha}$ act
on spin and orbital degrees of freedom, respectively. The basis takes
$(C_{s\uparrow},C_{s\downarrow},C_{p\uparrow},C_{p\downarrow})^{T}$.
In Eq.\ (\ref{eq:Hamiltonian}), the mass term $M(\mathbf{k})\equiv m+t_{z}\cos k_{z}+t_{\parallel}(\cos k_{x}+\cos k_{y})$
includes both an orbital energy difference $m$, and nearest-neighbor
hopping amplitudes $t_{z}$ and $t_{\parallel}$. The two $\lambda$
terms refer to spin-orbit coupling. With $t_{AM}=0$, $h({\bf k})$
describes a 3D TI classified by weak, strong, and trivial insulating
phases \citep{FuL07prb,fu2007topological,ZhangHJ09nature}. The term
$t_{AM}(\cos k_{x}-\cos k_{y})\rho_{z}\sigma_{z}$ represents a $d$-wave
altermagnetic order with field strength $t_{AM}$ \footnote{Note that the altermagnetic term starkly contrasts with the $(\cos k_{x}-\cos k_{y})\rho_{y}\sigma_{0}$
term in higher-order topological insulators \citep{Schindler18SA},
although both terms share the same symmetry. The $(\cos k_{x}-\cos k_{y})\rho_{y}\sigma_{0}$
term gaps all side Dirac cones whereas the altermagnetic term induces
anisotropic momentum shifts as we show below.}. It is similar to the altermagnetic order recently introduced in
2D quantum spin Hall insulators\textcolor{blue}{{} \citep{ma2024altermagnetic,gonzalez2025model,li2025floating,gonzalez2025spin}}.

\begin{figure}
\includegraphics[width=8.5cm]{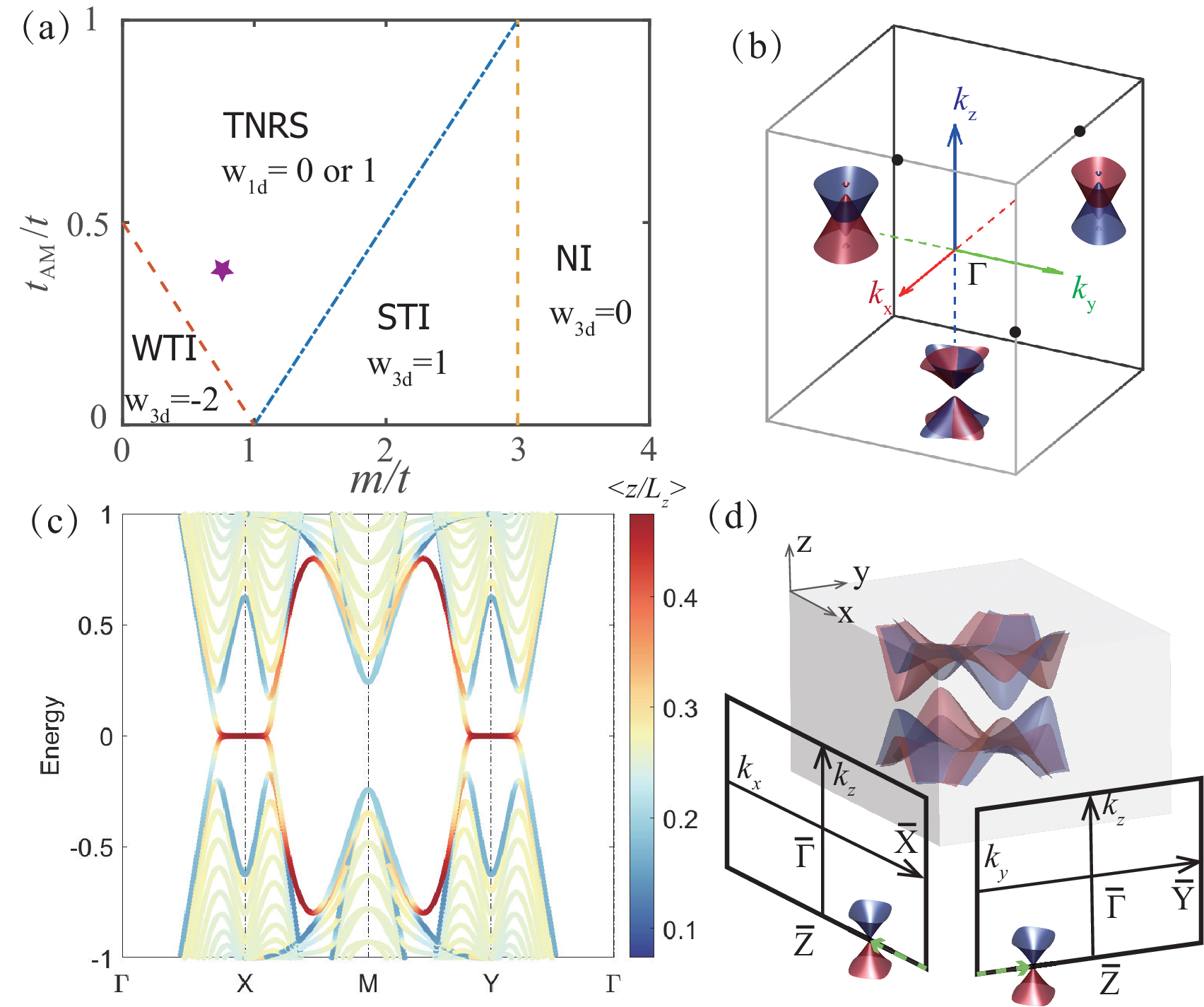}\caption{(a) Phase diagram of the AMTI in the parameter space of $m$ and $t_{AM}$.
The four different phases are weak topological insulators (WTIs),
strong topological insulators (STIs), normal insulators (NIs), and
topological nodal-ring semimetals (TNRS), respectively. (b) 3D Brillouin
zone and band structure for TNRS {[}purple hexagon in (a){]} near
the high-symmetry points $(\pi,0,\pi)$, $(0,\pi,\pi)$ and $(\pi,\pi,0)$
(marked by black dots). The plots show the band dispersion for $k_{z}=0$
or $\pi$. Red and blue colors denote the horizontal mirror eigenvalues
$\sigma_{h}=+i$ and $-i$, respectively. (c) Band structure of a
thin-film AMTI in the $xy$-plane with width $L_{z}=20$, plotted
along high-symmetry lines. Parameters: $t_{AM}=0.4$, $\lambda=1$,
and $m=0.8$. (d) Schematic of the AM-induced anisotropic shift of
surface Dirac cone on (010) and (100) side surfaces in the STI phase.
\label{fig:normal_state}}
\end{figure}

Before discussing the superconducting phase, it is instructive to
analyze the properties of the normal-state Hamiltonian $h(\mathbf{k})$
of the topological AMs. With $t_{AM}\neq0$, the altermagnetic order
profoundly influences the original properties of the TIs. Importantly,
it modifies the symmetry of the system. The altermagnetic order breaks
both time-reversal symmetry $T$ and four-fold rotation symmetry $C_{4z}$
individually, but preserves their combined operation $C_{4z}T$. This
reduced symmetry shift the system from the magnetic point group $4/mmm1'$
to $4'/m'm'm$, resulting in an anisotropic Fermi surface {[}see Fig.
\ref{fig:normal_state}(b){]} and momentum-dependent spin polarization
\citep{FuBo2025SM}. Additionally, the altermagnetic order affect
the topological properties and band structure significantly. The modified
phase diagram is presented in Fig. \ref{fig:normal_state}(a). The
anisotropic nature of the altermagnetic order makes it act differently
on the high-symmetry points (HSPs) of the Brillouin zone. Near the
$(\pi,0,\pi)$ and ($0,\pi,\pi$) points, it acts as a constant Zeeman
field $\sim2t_{AM}\rho_{z}\sigma_{z}$, lifting the spin degeneracy.
While near the $(\pi,\pi,0)$ point, it takes $d$-wave form $\sim\frac{t_{AM}}{2}(\delta k_{x}^{2}-\delta k_{y}^{2})\rho_{z}\sigma_{z}$
(where $\delta k_{i}$ is the momentum deviation from the HSP), preserving
band degeneracy at this point but splitting bands away from it. When
$2t_{AM}>|M({\bf k})-t|$, the band inversion near ($\pi,0,\pi$)
and $(0,\pi,\pi)$ creates a nodal structure in the $k_{z}=\pi$ plane,
yielding a topological nodal-ring semimetal phase {[}see Figs. \ref{fig:normal_state}(a)
and \ref{fig:normal_state}(b){]}. Its topology is captured by the
1D winding number and manifested by zero-energy states at boundaries
as shown in Fig. \ref{fig:normal_state}(c) \footnote{This 1D winding number is computed along $k_{z}$-direction for a
fixed in-plane momentum $(k_{x},k_{y})$. More details can be found
in the Ref. \citep{FuBo2025SM}.}. Due to the presence of chiral symmetry $\mathcal{C}=\rho_{y}$,
the gapped phases are characterized by the 3D winding number $w_{3D}$
\citep{ChiuCK16rmp}. This invariant distinguishes the topological
phases: $w_{3D}=1$ for STI, $w_{3D}=-2$ for WTI, and $w_{3D}=0$
for NI (see End Matter).

Notably, the anisotropic nature of AMs strongly affects the topological
surface states. For concreteness, we focus on the STI phase of the
phase diagram Fig. \ref{fig:normal_state}(a) in the following. A
low-energy expansion near the ($\pi,\pi,\pi$) point demonstrates
that the altermagnetic term takes the form $\sim t_{AM}(\delta k_{x}^{2}-\delta k_{y}^{2})\rho_{z}\sigma_{z}$.
On certain side surfaces, such as the (100) and (010) facets, the
altermagnetic order acts as a direction-dependent Zeeman field, shifting
the Dirac cones in an anisotropic way {[}see Fig. \ref{fig:normal_state}(d){]}.
For instance, the effective Hamiltonian for the (010) surface is given
by: $h_{010}=\lambda[(-\delta k_{x}+k_{AM})\sigma_{x}-\delta k_{z}\sigma_{y}]$
with $k_{AM}\equiv-\frac{t_{AM}}{t}\frac{(m-3t)}{\lambda}$ \citep{FuBo2025SM}.
In contrast, the projection of the AM term vanishes on (110), ($\bar{1}$10)
and (001) surfaces, leaving their Dirac cones intact. This is crucial
for the formation of BFSs and the realization of topological superconductivity,
as we discuss below.

\begin{figure}
\includegraphics[width=9cm]{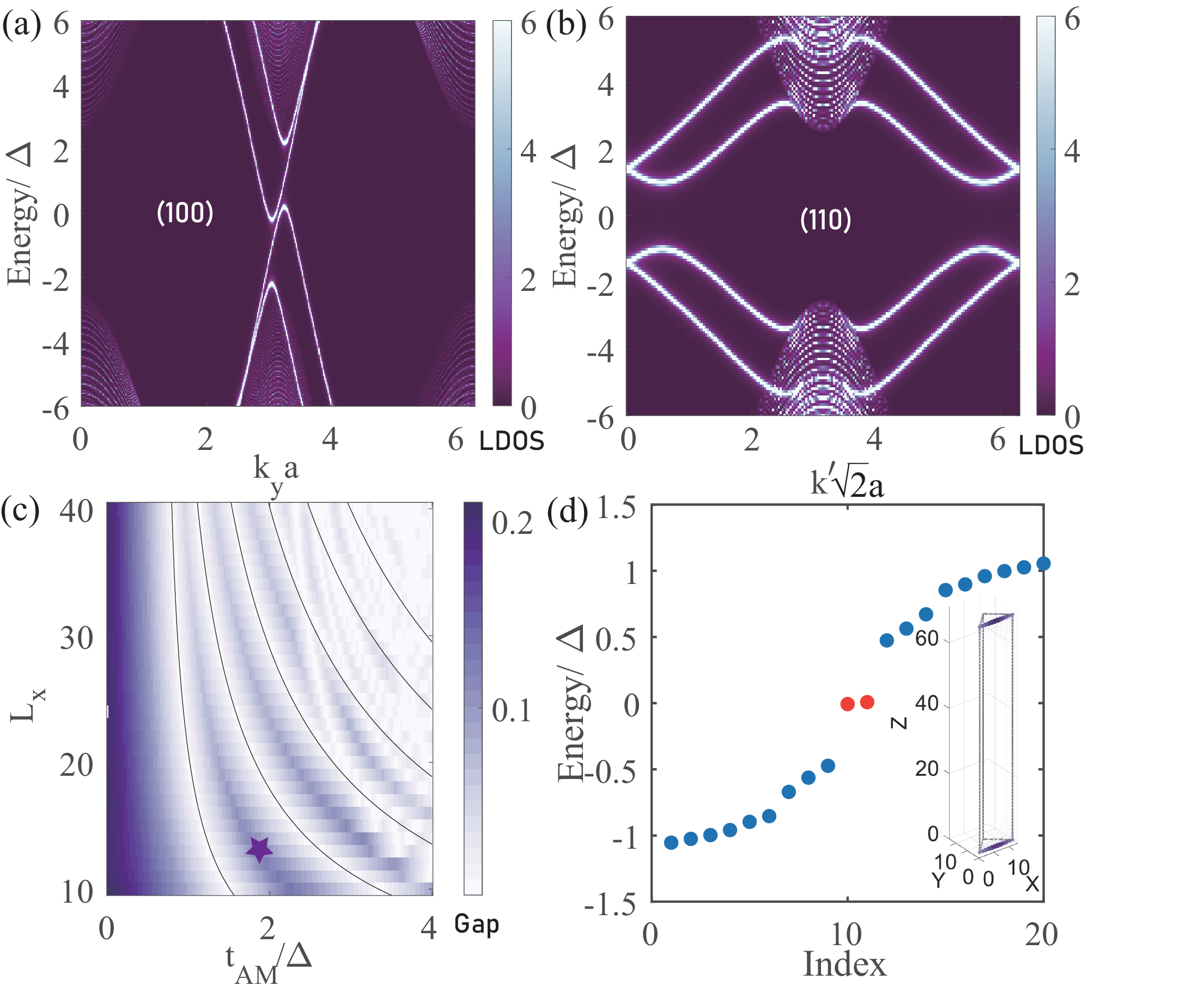}\caption{(a, b) Local density of states on (010) and (110) surfaces, respectively,
at $k_{z}=\pi$ calculated from the surface Green's function. (c)
Topological phase diagram for a triangular prism geometry with surfaces
$(110)$, ($1\bar{1}0$), and $(010)$. The color map represents the
energy gap obtained from numerical diagonalization of the quasi-1D
nanowire at $k_{z}=\pi$, plotted as a function of $L_{x}$ and $t_{AM}$.
Black curves indicate phase boundaries obtained from analytical solutions.
(d) Energy spectra for quasi-1D nanowire of length $L_{z}=80$. The
red dots indicate MZMs. The inset shows the spatial profile of the
MZMs. Other parameters are: $m=1.5$, $\mu=0.1$, $\lambda=1$, and
$\Delta=0.1$. \label{fig:edge-band-structure}}
\end{figure}

In the presence of superconductivity, the system possesses particle-hole
symmetry $\Xi=\tau_{y}\sigma_{y}K$ where $K$ is complex conjugation.
A finite chemical potential breaks chiral symmetry explicitly, placing
the system in the topologically trivial class D for 3D gapped systems.
However, the altermagnetic order can induce gapless bulk and surface
states, which host nontrivial topology. For the bulk states, the altermagnetic
spin splitting prohibits the formation of conventional spin-singlet
Cooper pairs between time-reversed states at $\mathbf{k}$ and $-\mathbf{k}$.
This leads to a superconducting state with a gapless quasiparticle
spectrum \footnote{This gapless BFS demonstrates diverse topological structures, such
as sphere and torus, when the chemical potential lies in the bulk
bands and the condition $t_{AM}>t_{c}^{b}=\frac{1}{2}\sqrt{\Delta^{2}+(\mu-|m-t|)^{2}}$
is satisfied (see Ref. \citep{FuBo2025SM}).}, manifested as bulk BFSs \citep{brydon2018bogoliubov,fukaya2025crossed,pal2024identifying,wu2025intra,setty2020topological,timm2021symmetry,oh2021using,yuan2022supercurrent,mo2025coexistence}.
Particularly, we are more interested in the surface bands, when the
chemical potential lies in the bulk band gap. With a superconducting
gap $\Delta,$ the surface spectrum for the (100) surface at $\delta k_{z}=0$
becomes: $E_{s,\zeta}^{100}=s\lambda k_{AM}+\zeta\sqrt{\Delta^{2}+(\mu+s\lambda\delta k_{y})^{2}}$,
where $s,\zeta=\pm1$. While for the (110) surface, its spectrum remains
intact as $E_{s,\zeta}^{110}=\zeta\sqrt{\Delta^{2}+(\mu+s\lambda\delta k^{\prime})^{2}}$.
The local density of states calculated for a semi-infinite system
\citep{smidstrup2017first} for these two cases are shown in Figs.
\ref{fig:edge-band-structure}(a) and \ref{fig:edge-band-structure}(b),
respectively. For the (100) facet, there exists a critical altermagnetic
strength $t_{c}^{s}=\frac{\Delta t}{|m-3t|}$, beyond which the surface
BFSs form {[}see Fig. \ref{fig:edge-band-structure}(a){]}. In contrast,
there is a direct superconducting gap for the (110) facet {[}see Fig.
\ref{fig:edge-band-structure}(b){]}. Both bulk and surface BFSs are
topologically protected \citep{FuBo2025SM}. For bulk BFSs, they are
associated with a $\mathbb{Z}_{2}$ invariant in presence of inversion
and particle-hole symmetry. The surface BFSs rely on the topology
of the parent AMTI characterized by a $\mathbb{Z}$ invariant. Since
the BFSs are topologically protected, they are robust against weak
disorder \citep{FuBo2025SM}.

\textit{\textcolor{blue}{Engineering topological superconductivity
via surface anisotropy with altermagnetic order.---}}Considering
the setup in Fig. \ref{Schematic_diagram}(b), we now demonstrate
how the anisotropic surface properties and associated BFS can be exploited
to obtain a topological superconducting phase hosting MZMs. The quasi-1D
nanowire is extended along $z$-direction. We use a triangular prism
geometry bounded by $(110)$, ($1\bar{1}0$), and $(010)$ crystallographic
facets, which breaks $C_{4z}T$ symmetry lifting degeneracies. To
investigate the topological properties of this system, we assume the
nanowire has a finite width $L_{x}$ in the (010) facet and infinite
length in $z$ direction. Henceforth, $k_{z}$ is a good quantum number.
Due to quantum confinement, the BFSs at surface $(100)$ evolve to
different energy bands as a function of $k_{z}$ within the superconducting
gap $\Delta$. This is because the BFSs on the (010) facet are bounded
by gapped surface states on the $(110)$ and ($1\bar{1}0$) facets
with a finite superconducting gap. The topological property of this
quasi-1D system is characterized by a $\mathbb{Z}_{2}$ invariant
in class D, enabled by the presence of particle-hole symmetry \citep{ChiuCK16rmp}.
The topological phase transition is determined by the gap closing
at $k_{z}=\pi$, which is tuned by both the size $L_{x}$ and strength
$t_{AM}$.

The gap-closing conditions for topological phase transitions can be
obtained by employing a low-energy effective theory of surface states
\citep{FuBo2025SM}, which yields

\begin{equation}
L_{x}/\xi=\frac{n\pi-\arctan\sqrt{(k_{AM}\xi)^{2}-1}}{\sqrt{(k_{AM}\xi)^{2}-1}},
\end{equation}
where $\xi=\lambda/\Delta$ is the superconducting coherence length,
$k_{AM}=-\frac{t_{AM}}{t}\frac{(M-3t)}{\lambda}$, and $n$ is an
integer taking $n=1,2,3,\cdots$. This equation determines the phase
boundaries between trivial and topological superconductors in the
parameter space $(t_{AM},L_{x})$, as shown by solid lines in Fig.
\ref{fig:edge-band-structure}(c). It is consistent with the result
obtained from Andreev scattering method \citep{papaj2021creating}.
We can directly compare the analytical results with numerical simulations.
The energy gap obtained from a tight-binding model is presented in
Fig. \ref{fig:edge-band-structure}(c). The quasi-1D nanowire exhibits
a series of gap closings and re-openings, indicating multiple topological
phase transitions. The excellent agreement between these analytical
results and numerics confirms the above analysis of phase transitions.

We further examine the topological nature of the system by showing
the appearance of MZMs driven by altermagnetic orders. Let us focus
on the first topological phase region between $n=1$ and $n=2$ with
a topological invariant $\mathbb{Z}_{2}=1$ \footnote{The $\mathbb{Z}_{2}$ invariant can be calculated following Kitaev's
formula as $(-1)^{\nu}=\mathrm{sgn}(\mathrm{Pf}[iH(k_{z}=0)]\mathrm{Pf}[iH(k_{z}=\pi)])$
from Pfaffian of the Hamiltonian at two points.}, as shown in Fig. \ref{fig:edge-band-structure}(c). We impose open
boundary conditions along $z$ direction and take a finite length
$L_{z}$. The energy spectra are plotted in Fig. \ref{fig:edge-band-structure}(d).
A pair of MZMs is found to reside at the band center, with a protecting
gap as large as the original superconducting gap $\Delta$. This large
gap provides advantages avoiding the fine tuning of parameters. The
corresponding wave functions are shown in the inset of Fig. \ref{fig:edge-band-structure}(d),
which indicates that the hosted MZMs localize at the two ends of quasi-1D
nanowire as expected.

\textit{\textcolor{blue}{Altermagnetic control of MZMs in a vortex
line.---}}Next, we consider the setup in Fig. \ref{Schematic_diagram}(c)
and demonstrate the effective control of MZMs in the vortex lines
by altermagnetic order. We can implement a $\pi$-flux vortex line
along $z$-direction, described by the order parameter $\Delta\tanh(r/\xi)e^{i\theta_{r}}$
where $r=\sqrt{x^{2}+y^{2}}$ and $\theta_{r}$ is the phase angle
and $\xi$ the coherence length. The vortex line breaks translational
symmetry in the $xy$-plane but keeps $k_{z}$ a good quantum number.
With a finite cross-section in the $xy$-plane, the system hosting
a vortex line can be treated as quasi-1D.

\begin{figure}
\includegraphics[width=8.5cm]{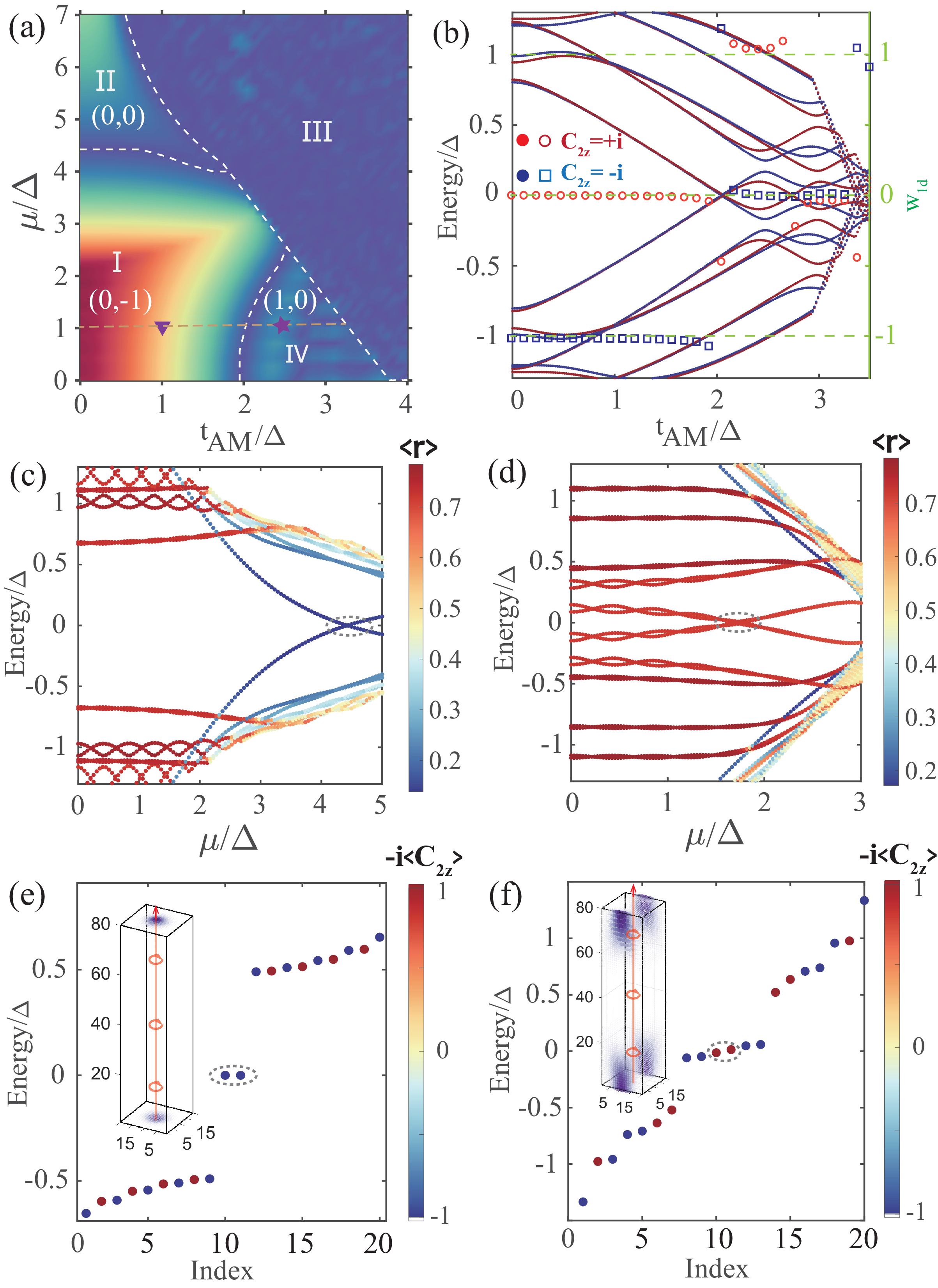}\caption{(a) Topological phase diagram in the parameter space of altermagnetic
strength $t_{AM}$ and chemical potential $\mu$ for a vortex line
along the $z$-axis. Roman numerals label distinct phases: (I) Majorana
vortex phase, (II) trivial vortex phase, (III) bulk nodal phase, and
(IV) surface nodal phase. The color scale represents the energy gap
near zero energy. (b) Energy spectrum $E(k_{z}=\pi)$ as a function
of $t_{AM}$ at a fixed chemical potential $\mu/\Delta=1$ {[}horizontal
dashed line in (a){]}. Red and blue lines denote states in $C_{2z}=+i$
and $-i$ eigensectors, with squares and circles marking the corresponding
topological winding numbers $(w_{+i},w_{-i})$. (c, d) Energy spectrum
$E(k_{z}=\pi)$ as a function of $\mu$ for a fixed altermagnetic
strength: In (c) $t_{AM}/\Delta=0.5$ and in (d) $t_{AM}/\Delta=2.3$.
The color of each data point indicates the spatial localization of
the corresponding state, quantified by its average distance from the
vortex core $\langle r\rangle$ (red: outer surface states; blue:
vortex core-bound states). (e, f) Energy spectrum with open boundary
conditions along $z$-direction ($L_{z}=80$) and spatial profile
of the corresponding MZMs (inset) for parameters marked in (a): (e)
$t_{AM}/\Delta=1$ (triangle), (f) $t_{AM}/\Delta=2.4$ (hexagon).
Parameters for all panels: $L_{x}=L_{y}=20$, $m=2.5$, $\lambda=1$,
and $\Delta=0.2$.\label{fig:vortex}}
\end{figure}

We obtain the energy spectrum of this quasi-1D system by diagonalizing
the tight-binding BdG Hamiltonian. The topological phase transition
is captured by the gap closing at $k_{z}=\pi$. The topological phase
diagram with corresponding gap values in parameter space $(t_{AM},\mu)$
is presented in Fig. \ref{fig:vortex}(a). The color scale represents
the band gap of the energy spectrum at $k_{z}=\pi$. This phase diagram
is divided into four distinct phase regions, labeled as (I) to (IV).
The topological properties of the gapped regions can be characterized
by a pair of winding numbers $\mathbb{Z}_{+i}\oplus\mathbb{Z}_{-i}\equiv(w_{+i},w_{-i})$,
which are defined as follows. Despite the breaking of time reversal
symmetry $T$ by the vortex line, the system retains a modified symmetry
resembling time reversal, i.e., the combined operation $\sigma_{v}T$,
where $\sigma_{v}$ denotes the vertical mirror reflection with respect
to the $xz$ or $yz$-plane. The chiral symmetry $\mathcal{C}_{v}$
can be further constructed from the particle-hole symmetry $\Xi$
and $\sigma_{v}T$, given by $\mathcal{C}_{v}=\tau_{z}\rho_{z}\sigma_{z}$
incorporating the mirror reflection with $\mathcal{C}_{v}^{2}=1$.
Furthermore, since $[\mathcal{C}_{v},C_{2z}]=0$, a winding number
can be defined within each eigensector of the two-fold rotation $C_{2z}$.
Consequently, the quasi-1D vortex system is characterized by a pair
of winding numbers $(w_{+i},w_{-i})$, where the subscript denotes
the $C_{2z}$ eigenvalue \citep{qin2019congcong,hu2022competing,hu2023topological,luo2025surface}.
These invariants determine the presence or absence of MZMs at the
two ends of the system.

The four phase regions shown in Fig. \ref{fig:vortex}(a) demonstrate
different topological properties and band gaps. Explicitly, they are
classified as: (I) Majorana vortex phase at small $\mu$ and small
$t_{AM}$, in which the vortex line is topological nontrivial and
hosts MZMs at its ends; (II) trivial vortex phase at large $\mu$
and small $t_{AM}$; (III) bulk nodal phase for $t_{AM}>t_{c}^{b}$
with $t_{c}^{b}=\frac{1}{2}\sqrt{\Delta^{2}+(\mu-|m-t|)^{2}}$, in
which the bulk bands develop BFSs; and (IV) surface nodal vortex phase
for $t_{c}^{b}>t_{AM}>t_{c}^{s}$. In this particular phase (IV),
the system has an oscillating band gap while hosting MZMs at the surface
boundaries. We mainly focus on the properties of nontrivial regions
(I) and (IV) in the following \footnote{Note that the phase region (I) also appears in superconducting vortices
of doped topological insulators \citep{hosur2011majorana}, while
the phase region (IV) is unique due to the presence of altermagnetic
order}.

The vortex phase transition can be induced by the altermagnetic order
strength $t_{AM}$ and chemical potential $\mu$. At fixed chemical
potential $\mu/\Delta=1$, the low-energy spectrum of the quasi-1D
system as a function of $t_{AM}$ is shown in Fig. \ref{fig:vortex}(b)
{[}horizontal dashed line in Fig. \ref{fig:vortex}(a){]}. The spectra
for $C_{2z}=+i$ and $-i$ eigensectors are depicted in red and blue,
respectively. Note that eigenstates from different $C_{2z}$ sectors
will not hybridize due to the symmetry protection. As $t_{AM}$ increases,
the spectral gaps in both sectors close near $t_{AM}/\Delta=2\simeq t_{c}^{b}$,
indicating a topological phase transition. This transition is also
confirmed by the jump of the winding numbers $(w_{+i},w_{-i})$ from
$(0,-1)$ to $(1,0)$, annotated by square and circle symbols, respectively.
To understand the vortex phase transitions and surface phase transition
of BFSs at the outer surfaces, we plot the spectra of the quasi-1D
system at $k_{z}=\pi$ as a function of $\mu$ for different $t_{AM}$
in Figs. \ref{fig:vortex}(c) and \ref{fig:vortex}(d), respectively.

In the subcritical regime with $t_{AM}<t_{c}^{b}$ {[}see Fig. \ref{fig:vortex}
(c){]}, the primary low-energy physics and topological phase transitions
are governed by the vortex line. At small $\mu$, the vortex line
is topological nontrivial and hosts a pair of MZMs at its two ends
\citep{hosur2011majorana,yan2020vortex,ghorashi2020vortex}. This
topological characterization persists as $\mu$ increases into the
bulk states until a topological phase transition occurs at the point
marked by the gray circle in Fig. \ref{fig:vortex}(c). Intriguingly,
discrete low-energy states appear within the superconducting gap in
this phase (at small $\mu$), which are bound to the side surfaces
rather than the vortex line. In stark contrast, in the supercritical
regime with $t_{AM}>t_{c}^{b}$ {[}see Fig. \ref{fig:vortex} (d){]},
the dominant physics and topological phase transitions shift to the
outer side surfaces. Due to the quantum confinement in the $xy$-plane,
the surface BFSs are quantized to discrete energy levels within the
superconducting gap $\Delta$. The introduction of a $\pi$-flux vortex
line breaks the $C_{4z}T$ symmetry and lifts the double degeneracy
of energy bands. This lifting of degeneracy ultimately endows a novel
topological phase transition {[}gray circle in Fig. \ref{fig:vortex}(d){]}
of BFSs at the outer side surface as changing $\mu$. Note that there
is no band crossing at the vortex line in this case.

Finally, we present the energy spectra and corresponding probability
density of MZMs under open boundary conditions for the two distinct
topological phases. For the phase region (I) with winding numbers
$(w_{+i},w_{-i})=(0,-1)$, indicated by a triangle in Fig. \ref{fig:vortex}(a),
the MZMs are localized at the vortex line {[}see Fig. \ref{fig:vortex}(e){]},
consistent with the Fu-Kane proposal \citep{FuL08prl,hosur2011majorana}.
In contrast, for the phase region (IV) $(w_{+i},w_{-i})=(1,0)$, represented
by the hexagon in Fig. \ref{fig:vortex}(a), the MZMs are localized
at the outer surface {[}see Fig. \ref{fig:vortex}(f){]} \footnote{Note that the eigenstates from different $C_{2z}$ sectors presented
in Fig. \ref{fig:vortex}(f) will not hybridize. Thus the MZMs are
still be protected by a superconducting energy gap as large as $\Delta$.}. It is interesting to note that the $C_{2z}$ eigenvalues corresponding
to MZMs of these two cases are also switched. This result demonstrates
that the altermagnetic order drives a topological phase transition
which relocalizes MZMs from the vortex line core to the outer side
surfaces.

\textit{\textcolor{blue}{Discussion and conclusions.---}}We propose
a fundamentally new way to realize topological superconductivity utilizing
the unique interplay between $d$-wave AMs, topology, and $s$-wave
superconductivity. Our proposal is based on particular features of
this combination: (i) Crystal-facet-dependent anisotropy: The surface
BFSs at different physical boundaries are anisotropic. Thus, quantum
confinement decisively affects the topological phase. This mechanism
is substantially different from previous proposals \citep{Ghorashi24PRL,li2024realizing,li2024creation}
on the realization of MZMs and topological superconductivity in AM
based systems \footnote{Reference \citep{Ghorashi24PRL} suggests to substitute the uniform
magnetic field or ferromagnetic order in previous proposals, such
as 1D semiconductor nanowires and 2D chiral superconductors, by altermagnetic
order. References \citep{li2024creation,li2024realizing} utilize
altermagnetic order to create higher-order topology and Majorana corner
modes in 2D systems.}. (ii) Stray field free and zero net magnetization: There is no need
for external magnetic fields, which excludes detrimental stray field
that jeopardizes superconductivity and scalability. (iii) Vortex-line
based Majorana manipulation: We demonstrate how MZMs move from the
vortex line to the side surfaces driven by the altermagnetic order
via a topological phase transition.

Our proposal could be realized in a heterostructure consisting of
AM-TI-SC, utilizing the proximity effect of both altermagnets \citep{zhu2025altermagnetic}
and superconductors. There are several available AM materials including
$\mathrm{KV_{2}Se_{2}O}$ \citep{Jiang25NP}, $\mathrm{Rb}_{1-\delta}\mathrm{V_{2}Te_{2}O}$
\citep{ZhangF25NP}, and $\mathrm{Mn_{5}Si_{3}}$ \citep{Reichlova24nc,HanL25sa},
that could be brought in proximity to 3D TIs and $s$-wave superconductors.\textcolor{blue}{{}
}Both proposed topological phases rely on the intrinsic hierarchy
of energy scales $t_{c}^{s}<t_{AM}<t_{c}^{b}$, where $t_{c}^{s}\propto\Delta$
is pairing-limited and $t_{c}^{b}\sim t$ is bandwidth-limited\textcolor{blue}{{}
\citep{FuBo2025SM}}. Alternatively, our proposal could also be realized
based on candidate material $\mathrm{EuIn_{2}As_{2}}$ in proximity
to an $s$-wave superconductor \citep{FuBo2025SM}. This material
exhibits complex antiferromagnetic structures \citep{XuY19prl,Sato20prr,cuono2023ab,Riberolles21NC,pari2025nonrelativistic}
and possible $g$-wave AMTI could be supported in the afmc phase \textcolor{blue}{\citep{FuBo2025SM}.} We
expect that our predictions, including the facet-dependent superconducting
gap and the spatial shift of MZMs, could be observed via scanning
tunneling microscopy/spectroscopy (STM/STS) \citep{XuJP-15prl} or
angle-resolved photoemission spectroscopy (ARPES) \citep{gao2024arpes}.

\textit{\textcolor{blue}{Acknowledgments.---}}We thank J. Li for
helpful discussions. B.F. was financially supported by National Natural
Science Foundation of China (Grants No.12504049), Guangdong Province
Introduced Innovative R\&D Team Program (Grant No. 2023QN10X136),
Guangdong Basic and Applied Basic Research Foundation No. 2024A1515010430
and 2023A1515140008), Quantum Science Center of Guangdong-Hong Kong-Macao
Greater Bay Area (Grant No. GDZX230005). C.A.L. and B.T. were supported
by the DFG (SFB 1170 ToCoTronics), and the Würzburg-Dresden Cluster
of Excellence ctd.qmat, EXC 2147 (Project-Id 390858490).

\bibliographystyle{apsrev4-2}
\bibliography{Refsdata}

\section*{End Matter}

\textit{Symmetry analysis of the AMTIs.---}The introduction of altermagnetic
order fundamentally reconstructs the symmetry landscape of the system.
The parent state of TIs belongs to the magnetic point group 4/mmm1',
which includes all unitary crystalline symmetries of the $D_{4h}$
point group and their products with time-reversal symmetry $T$.

The specific form of the altermagnetic order, which couples as a momentum-dependent
Zeeman field with a $d$-wave symmetry, breaks time-reversal symmetry
$T$. However, due to its compensated and non-ferromagnetic nature,
it preserves a subset of symmetries that combine $T$ with broken
unitary operations. The resulting system is described by the magnetic
point group $4'/m'm'm$. The fate of each symmetry operation under
the introduction of altermagnetism is summarized in Table I and analyzed
below.

\textit{Topological invariants of nodal-ring semimetals and superconducting
phases.---}The topological characterization of the system evolves
significantly between the normal and superconducting states, and is
further refined in the presence of defects such as vortices. In the
normal state, the Hamiltonian of AMTIs possesses a chiral symmetry
given by $\mathcal{C}=\rho_{y}$, satisfying $\mathcal{C}^{2}=1$.
When the spectrum is fully gapped, the system is classified by a three-dimensional
winding number:

\begin{equation}
w_{3D}=\int\frac{d^{3}\mathbf{k}}{48\pi^{2}}\epsilon_{abc}\mathrm{Tr}[\mathcal{C}\mathcal{H}^{-1}\partial_{k_{a}}\mathcal{H}\mathcal{H}^{-1}\partial_{k_{b}}\mathcal{H}\mathcal{H}^{-1}\partial_{k_{c}}\mathcal{H}].
\end{equation}
However, in the presence of gapless nodal rings, this invariant is
ill-defined. The topology of the nodal structure itself is instead
captured by a 1D winding number computed on a path along $z$-direction
in momentum space. For a ring extending in the $k_{x}-k_{y}$ plane,
the invariant for a given $(k_{x},k_{y})$ is
\begin{equation}
w_{1D}(k_{x},k_{y})=\int\frac{dk_{z}}{4\pi i}\mathrm{Tr}[\mathcal{C}\mathcal{H}^{-1}\partial_{k_{z}}\mathcal{H}].
\end{equation}

\begin{table}
\begin{tabular}{|c|c|c|c|c|c|c|c|c|c|}
\hline 
$E$ & $2C_{4z}$ & $C_{2z}$ & $2C_{2}^{\prime}$ & $2C_{2}^{\prime\prime}$ & $\mathcal{I}$ & $2S_{4}$ & $\sigma_{h}$ & $2\sigma_{v}$ & $2\sigma_{d}$\tabularnewline
\hline 
\CheckmarkBold{} & \ding{56} & \CheckmarkBold{} & \ding{56} & \CheckmarkBold{} & \CheckmarkBold{} & \ding{56} & \CheckmarkBold{} & \ding{56} & \CheckmarkBold{}\tabularnewline
\hline 
$T$ & $2TC_{4z}$ & $TC_{2z}$ & $2TC_{2}^{\prime}$ & $2TC_{2}^{\prime\prime}$ & $\mathcal{I}T$ & $2TS_{4}$ & $T\sigma_{h}$ & $2T\sigma_{v}$ & $2T\sigma_{d}$\tabularnewline
\hline 
\ding{56} & \CheckmarkBold{} & \ding{56} & \CheckmarkBold{} & \ding{56} & \ding{56} & \CheckmarkBold{} & \ding{56} & \CheckmarkBold{} & \ding{56}\tabularnewline
\hline 
\end{tabular}

\caption{Symmetry analysis of the $4'/m'm'm$ magnetic point group of the topological
AMs. The table classifies all crystalline symmetries as either preserved
or broken by the altermagnetic order. A checkmark (\CheckmarkBold )
indicates a symmetry that is preserved, while a cross (\ding{56})
indicates a symmetry that is broken. Symmetry operations are defined
as: $E$, identity; $C_{nz}$, $n$-fold rotation about $z$-axis;
$C_{2}^{\prime}/C_{2}^{\prime\prime}$ , two-fold rotations about
in-plane axes; $\mathcal{I}$, inversion; $S_{4}$, four-fold rotoinversion;
$\sigma_{h}/\sigma_{v}/\sigma_{d}$, mirror planes; $T$, time-reversal.
Prefixes indicate the number of symmetry-equivalent operations.}
\end{table}

Introducing superconductivity and a finite chemical potential generally
breaks the chiral symmetry $\mathcal{C},$ precluding the use of the
above invariants. A topological description can be recovered by considering
a quasi-1D geometry with open boundary conditions in the $xy$ plane
and periodic boundary conditions along $z$ direction. In this geometry,
the system respects a combined anti-unitary symmetry $\sigma_{v}T$,
where $\sigma_{v}$ is a mirror reflection with respect to the $xz$
or $yz$ plane. This symmetry allows for the construction of a new
chiral operator for the BdG Hamiltonian from the particle-hole symmetry
$\Xi$ and $\sigma_{v}T$, given by $\mathcal{C}_{1D}=\tau_{z}\rho_{z}\sigma_{z}$
incorporates the site mirror reflection with $\mathcal{C}_{1D}^{2}=1$.
The topology along the remaining periodic direction ($k_{z}$) is
then characterized by the winding number
\begin{equation}
w=i\int\frac{dk_{z}}{4\pi}\mathrm{Tr}[\mathcal{C}_{1D}\mathcal{H}_{\mathrm{BdG}}^{-1}(k_{z})\partial_{k_{z}}\mathcal{H}_{\mathrm{BdG}}(k_{z})].
\end{equation}
This framework can be extended to systems containing a vortex line
along $z$. While the vortex simultaneously breaks time-reversal symmetry
$T$ and mirror symmetry $\sigma_{v}$ individually, it preserves
their combination $\sigma_{v}T$. Consequently, the chiral symmetry
$\mathcal{C}_{1D}$ remains intact. The system additionally possesses
a two-fold rotational symmetry $C_{2z}=i\tau_{z}\sigma_{z}$ about
the $z$-axis. Given that $[\mathcal{C}_{1D},C_{2z}]=0$ and $[\Xi,C_{2z}]=0$,
the Hamiltonian can be block-diagonalized into the two eigensectors
($s=\pm$) of $C_{2z}$. A topological winding number can be defined
within each sector using the projector $\mathcal{P}_{s}=(1+siC_{2z})/2$:
\begin{equation}
w_{s}=i\int\frac{dk_{z}}{4\pi}\mathrm{Tr}[\mathcal{P}_{s}\mathcal{C}_{1D}\mathcal{H}_{\mathrm{BdG}}^{-1}(k_{z})\partial_{k_{z}}\mathcal{H}_{\mathrm{BdG}}(k_{z})].
\end{equation}
The total winding number is given by the sum over both sectors, $w=w_{+}+w_{-}$.
A nonzero $w_{s}$ in a given sector implies the existence of topological
protected vortex-bound states within that specific symmetry channel.

\clearpage
\onecolumngrid

\begin{center}
{\large\bfseries Supplemental material for
``Altermagnetism Induced Bogoliubov Fermi Surfaces Form Topological
Superconductivity''\par}
\vspace{0.7em}
Bo Fu$^{1}$, Chang-An Li$^{2,3}$, and Björn Trauzettel$^{4,5}$\par
\vspace{0.4em}
{\small
$^{1}$School of Sciences, Great Bay University, Dongguan, China\\
$^{2}$Hefei National Laboratory, Hefei, Anhui 230088, China\\
$^{3}$School of Emerging Technology, University of Science and Technology of China,
Hefei, Anhui 230026, China\\
$^{4}$Institute for Theoretical Physics and Astrophysics,
University of Würzburg, 97074 Würzburg, Germany\\
$^{5}$Würzburg-Dresden Cluster of Excellence ctd.qmat, Germany
}
\end{center}
\vspace{0.6em}

\begin{quote}
\small
This Supplemental Material provides detailed analytical and numerical
support for the main text, structured into four sections: In Sec.
S1, we present a comprehensive analysis of the Fermi surface topology
in an altermagnetic topological insulator and the consequent formation
of symmetry-protected Bogoliubov Fermi Surfaces (BFSs) upon the introduction
of $s$-wave superconductivity and disorder. In Sec. S2, we focus
on the surface spectrum. Using a $\mathbf{k}\cdot\mathbf{p}$ model
and open boundary conditions, we derive the effective Hamiltonian
for the topological surface states. We show that the altermagnetic
term acts as a momentum shift, and determine the critical strength
for the emergence of nodal surfaces in the presence of superconductivity.
In Sec. S3, we estimate relevant energy scales and consider the experimental
feasibility of our proposals. In Sec. S4, we investigate the topological
phase transition in a prism nanowire geometry. We solve for the bound
states and derive the exact phase boundary in the parameter space
of altermagnetic strength and system size. In Sec. S5, we extend our
analysis to vortex line along $x$-direction, demonstrating how the
altermagnetic order induces a characteristic spatial shift of Majorana
zero modes (MZMs) away from the vortex core. Finally, in Sec. S6,
we discuss possible g-wave altermagnetism topological insulator phase
in EuIn\textsubscript{2}As\textsubscript{2} and the realization of
topological superconductivity.
\end{quote}
\vspace{0.6em}


\renewcommand{\thefigure}{S\arabic{figure}}
\renewcommand{\thetable}{S\arabic{table}}
\renewcommand{\theequation}{S\arabic{equation}}
\renewcommand{\thesection}{S\arabic{section}}
\setcounter{figure}{0}  
\setcounter{equation}{0} 
\numberwithin{equation}{section}\tableofcontents{}

\section{Bogoliubov Fermi surfaces in bulk spectrum of a superconducting topological
altermagnet}

\begin{figure}
\includegraphics[width=10cm]{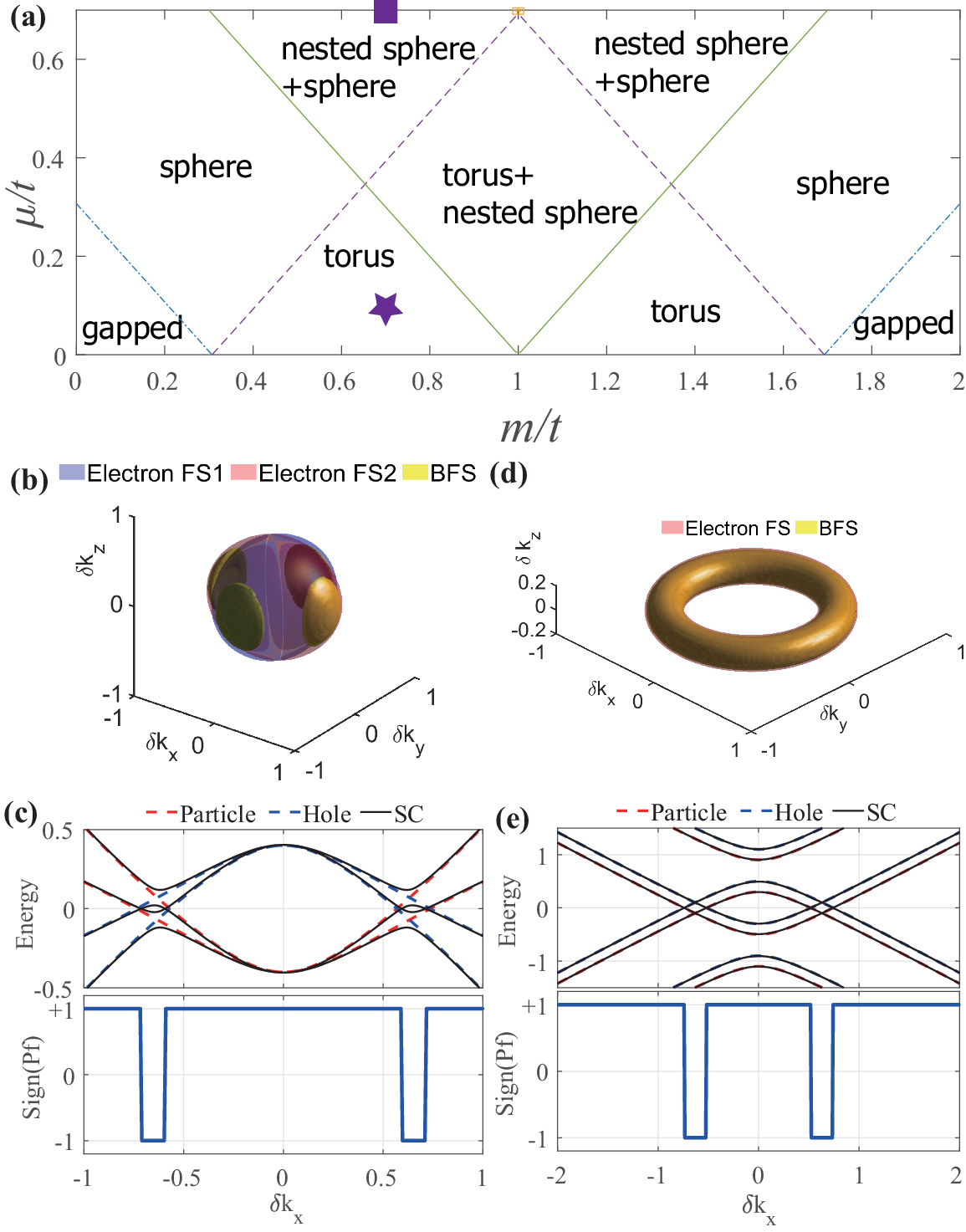}\caption{(a) Fermi surface topology of the altermagnetic topological insulators
as a function of the mass parameter $m$ and chemical potential $\mu$.
(b) Nested Fermi sphere geometry in the normal state (blue and red
surfaces) and the resulting 3D BFS (yellow surfaces) centered at ($\pi,\pi,0$)
for $\mu=0.7$ {[}purple hexagon in (a){]}. (c) Energy dispersion
and Pfaffian sign along $k_{x}$ in the vicinity of ($\pi,\pi,0$).
Red and blue dashed curves represent the particle and hole bands in
normal state, respectively. Black solid curves indicate the Bogoliubov
quasiparticle spectrum. (d, e) Corresponding plots near $(0,\pi,\pi)$
for $\mu=0.1${[}purple square in (a){]}, showing a torus Fermi surface
and its associated BFS. Other parameters are $t_{AM}=0.35$, $m=0.7$,
$\Delta=0.05$. \label{fig:Phases-in-chiral}}
\end{figure}

In this section, we present the topological properties of Bogoliubov
Fermi surfaces (BFSs) in the bulk spectrum. Let us start with the
normal states phase. Figure \ref{fig:Phases-in-chiral}(a) presents
the comprehensive Fermi surface topology in altermagnetic topological
insulator as a function of the mass parameter $m$ and chemical potential
$\mu$, for a fixed altermagnetic field strength $t_{AM}=0.35$. These
Fermi surfaces exhibit diverse topological configurations in momentum
space, including: (i) spherical surfaces centered at high-symmetry
points (HSPs), (ii) torus structures with non-trivial genus, and (iii)
nested spherical shells - with the specific morphology determined
by the $(m,\mu)$ parameters.

In the presence of superconducting pairing, the altermagnetic spin
splitting plays a crucial role by prohibiting the formation of conventional
spin-singlet Cooper pairs between time-reversed states at momenta
$\mathbf{k}$ and $-\mathbf{k}$. Consequently, due to the diverse
topology of the Fermi surface in a altermagnetic topological insulator,
the system stabilizes various types of BFSs. As shown in Figs. \ref{fig:Phases-in-chiral}(b)
and (d), we plot the BFSs in the vicinity of HSPs for two representative
cases, denoted by the purple square and hexagon in the phase diagram
of Fig. \ref{fig:Phases-in-chiral}(a). In Fig. \ref{fig:Phases-in-chiral}(b),
the normal-state Fermi surface consists of nested spheres, a result
of the altermagnetic spin splitting. The superconducting pairing can
only fully gap the states in the vicinity of the line defined by $\delta k_{x}=\pm\delta k_{y}$,
where the two Fermi surfaces cross. Here, $\delta k_{i}$ represents
the momentum deviation from the HSPs. Figure \ref{fig:Phases-in-chiral}(c)
shows a comparison of the band structure along the $k_{x}$ direction:
the particle (red dashed line) and hole bands (blue dashed line) in
the normal state, and the Bogoliubov quasiparticle spectrum (black
solid lines) in the superconducting state. This verifies that the
superconducting pairing cannot open a gap where the altermagnetic
spin splitting has separated the bands. In Fig. \ref{fig:Phases-in-chiral}(d),
the Fermi surface (red) exhibits a torus structure. In the presence
of superconducting pairing, the resulting BFS preserves this toroidal
geometry which is verified by the detailed band calculations in Fig.
\ref{fig:Phases-in-chiral}(e).

The microscopic origin of these BFSs can be understood through an
effective single-band description valid near the normal-state Fermi
surface. This description remains accurate under weak-coupling conditions
where the pairing potential \textgreek{\textDelta} is much smaller
than the direct band gap to other bands. The Bogoliubov-de Gennes
(BdG) Hamiltonian in the Nambu basis is:
\begin{equation}
\mathcal{H}_{\mathrm{BdG}}(\mathbf{k})=\left(\begin{array}{cc}
\mathcal{H}_{e}(\mathbf{k}) & \Delta\\
\Delta & \mathcal{H}_{h}(\mathbf{k})
\end{array}\right),
\end{equation}
where the hole block is related to the electron block by particle-hole
symmetry: $\mathcal{H}_{h}(\mathbf{k})=-\sigma_{y}\mathcal{H}_{e}^{*}(-\mathbf{k})\sigma_{y}$.
We use the HSP $(0,\pi,\pi)$ from Fig. \ref{fig:Phases-in-chiral}(d)
as an example. Expanding the electron Hamiltonian $\mathcal{H}_{e}(\mathbf{k})$
around this point, we obtain:
\begin{align}
\mathcal{H}_{e}(\mathbf{k}) & =(m-t)\rho_{z}\sigma_{0}+\lambda\rho_{x}(k_{x}\sigma_{x}-k_{y}\sigma_{y}-k_{z}\sigma_{z})\nonumber \\
 & +2t_{AM}\rho_{z}\sigma_{z}-\mu\rho_{0}\sigma_{0},
\end{align}
where $\rho_{i}$ and $\sigma_{i}$ are Pauli matrices acting on orbital
and spin space, respectively. This Hamiltonian is diagonalized as:
$\mathcal{H}_{e}(\mathbf{k})\Psi_{s,\zeta}^{e}(\mathbf{k})=E_{s\zeta}^{e}(\mathbf{k})\Psi_{s,\zeta}^{e}(\mathbf{k})$,
with the energy eigenvalues given by:
\begin{equation}
E_{s\zeta}^{e}(\mathbf{k})=s\sqrt{(\sqrt{(m-t)^{2}+\lambda^{2}k_{\perp}^{2}}+2\zeta t_{AM})^{2}+\lambda^{2}k_{z}^{2}},
\end{equation}
where $s,\zeta=\pm1$ and $k_{\perp}=\sqrt{k_{x}^{2}+k_{y}^{2}}$.
A unitary transformation $U=(\Psi_{++}^{e},\Psi_{+-}^{e},\Psi_{--}^{e},\Psi_{-+}^{e})$
diagonalizes the Hamiltonian: $\mathcal{H}_{e}(\mathbf{k})$ is diagonalized
as
\begin{equation}
U^{\dagger}\mathcal{H}_{e}(\mathbf{k})U=\mathrm{diag}(E_{++}^{e}(\mathbf{k}),E_{+-}^{e}(\mathbf{k}),E_{--}^{e}(\mathbf{k}),E_{-+}^{e}(\mathbf{k})).
\end{equation}
For the hole sector, the eigenvalues and eigenvectors are related
to their electronic counterparts by the particle-hole transformation:
$\mathcal{H}_{h}(\mathbf{k})\Psi_{s\zeta}^{h}(\mathbf{k})=E_{s\zeta}^{h}(\mathbf{k})\Psi_{s\zeta}^{h}(\mathbf{k})$
where $E_{s\zeta}^{h}(\mathbf{k})=-E_{s\zeta}^{e}(-\mathbf{k})$ and
$\Psi_{s\zeta}^{h}(\mathbf{k})=\sigma_{y}\Psi_{s\zeta}^{e*}(-\mathbf{k})$
are related to the electronic states via the particle-hole transformation.
Transforming into the eigenbasis of the normal state, the pairing
term undergoes the same unitary transformation. A direct calculation
shows that the intraband pairing matrix element vanishes: $\langle\Psi_{s\zeta}^{e}(\mathbf{k})|\Delta|\Psi_{s\zeta}^{h}(\mathbf{k})\rangle=0$.
This vanishing matrix element occurs because the torus-shaped Fermi
surface cannot support conventional spin-singlet superconducting states
with zero-momentum Cooper pairing. Consequently, the resulting BFSs
precisely inherit the topology of the normal-state Fermi surface,
preserving its toroidal geometry in the superconducting state. In
contrast, for the nested Fermi surface case in Fig. \ref{fig:Phases-in-chiral}(b),
conventional pairing can occur in the vicinity where the Fermi surfaces
intersect, allowing the superconducting gap to open in those regions.

The topological protection and classification of these BFSs are governed
by two fundamental symmetries, i.e., parity symmetry ($\mathcal{I}$)
and particle-hole ($\Xi$) symmetry. The combined $\mathcal{I}\Xi$
symmetry enforces a $\mathbb{Z}_{2}$ topological classification of
the system, where the invariant $\nu=sgn(P(\mathbf{k}_{+}))sgn(P(\mathbf{k}_{-}))$
is calculated via the relative sign of the Pfaffian at representative
momenta $\mathbf{k}_{\pm}$ on the two sides across the BFSs \citep{brydon2018bogoliubov}.
A value of $\nu=-1$ indicates a sign change of the Pfaffian, serving
as a definitive topological invariant that diagnoses the presence
of a stable, symmetry-protected BFS. As shown in the bottom panels
of Figs. \ref{fig:Phases-in-chiral}(c) and \ref{fig:Phases-in-chiral}(e),
our calculations of the Pfaffian sign along $k_{x}$ for both cases
confirm that it changes sign precisely where the BFSs appear.

We now derive the condition for BFS formation for arbitrary pairing
strength $\Delta$. For $k_{z}=0$, the Hamiltonian block-diagonalizes
into sectors $H_{+i}\oplus H_{-i}$, corresponding to eigenvalues
$\pm i$ under the horizontal mirror operator $\sigma_{h}.$ The energy
spectrum for an arbitrary pairing potential $\Delta$ is given by
\begin{equation}
E_{\chi}=2\chi t_{AM}\pm\sqrt{\Delta^{2}+(\mu\pm\sqrt{\lambda^{2}k_{\perp}^{2}+(m-t)^{2}})^{2}},
\end{equation}
where $\chi=\pm1$ is the sector index. Analysis of this spectrum
reveals that bulk nodal crossings occur when the altermagnetic term
exceeds a critical threshold, $t_{AM}>t_{c}^{b}$, where 
\begin{equation}
t_{c}^{b}\equiv\frac{1}{2}\sqrt{\Delta^{2}+(\mu-|m-t|)^{2}}.\label{eq:bulk_critical_strength}
\end{equation}
Furthermore, crystalline symmetry guarantees that an equivalent nodal
structure must appear at the $(\pi,0,\pi)$ point in the Brillouin
zone.

\begin{figure}
\includegraphics[width=10cm]{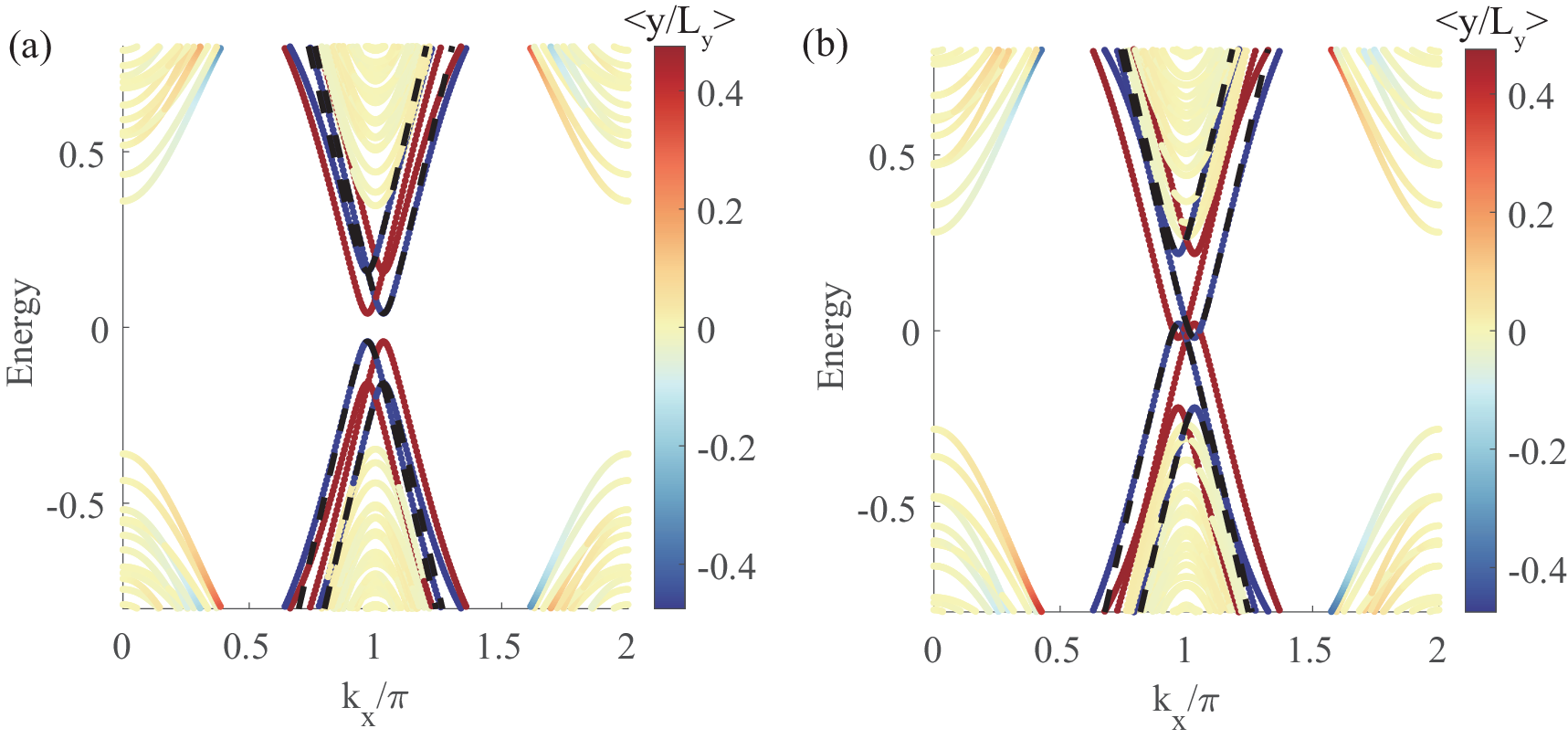}\caption{Band structure $E(k_{x},k_{z})$ at $k_{z}=\pi$ for a system with
open boundaries along the $y$-direction, computed for a system size
$L_{y}=20$. (a) Subcritical altermagnetic strength $t_{AM}=0.04$
($t_{AM}<t_{c}^{s}$). (b) Supercritical altermagnetic strength: $t_{AM}=0.08$
($t_{AM}>t_{c}^{s}$). The color gradient represents the spatial localization
of states on the two open edges. Black dashed lines show the analytical
solution from Eq. (\ref{eq:edge_state_with_SC-1}). Parameters: $m=1.5$,
chemical potential $\mu=0.1$, superconducting gap $\Delta=0.1$.
\label{fig:surface_nodal_transition}}
\end{figure}

\section{Bogoliubov Fermi surfaces in the surface spectrum of a superconducting
altermagnet topological insulator}

In this section, we present the formation of BFSs of the surface states
at physical boundaries. For concrete, we take parameters $m=1.5$
and $t_{AM}<0.25$, for which the the system resides in a strong topological
insulator phase characterized by a 3D winding number $w_{3d}=-1$.
To obtain the edge state spectrum, we employ the $\mathbf{k\cdot p}$
model near $(\pi,\pi,\pi)$ point. Imposing open boundary conditions
along the $y$ direction yields a Hamiltonian $\mathcal{H}=\mathcal{H}_{0}(-i\partial_{y})+\mathcal{H}_{1}$,
where
\begin{align}
\mathcal{H}_{0}(-i\partial_{y}) & =[(m-3t)-\frac{t}{2}\partial_{y}^{2}]\rho_{z}\sigma_{0}+i\lambda\partial_{y}\rho_{x}\sigma_{y},\\
\mathcal{H}_{1} & =-\lambda k_{x}\rho_{x}\sigma_{x}-\lambda k_{z}\rho_{x}\sigma_{z}+\frac{1}{2}t_{AM}\partial_{y}^{2}\rho_{z}\sigma_{z}.
\end{align}
The condition $(m-3t)t<0$ is necessary to guarantee the existence
of edge solutions. We first seek zero-energy solutions of $\mathcal{H}_{0}$
by solving $\mathcal{H}_{0}\Psi=0$. Using the ansatz $\Psi=\phi e^{\xi y}$,
we obtain:
\begin{equation}
\{[m-3t-\frac{t}{2}\xi^{2}]-\lambda\xi\rho_{y}\sigma_{y}\}\phi=0.
\end{equation}
The eigenstates of $\rho_{y}\sigma_{y}$ satisfy $\rho_{y}\sigma_{y}\phi_{\zeta}=\zeta\phi_{\zeta}$,
which leads to the characteristic equation:
\begin{equation}
m-3t-\frac{t}{2}\xi^{2}-\zeta\lambda\xi=0.
\end{equation}
Solving for $\xi$ yields:
\begin{equation}
\xi_{\zeta,\pm}=\frac{-\zeta\lambda\pm\sqrt{\lambda^{2}+2(m-3t)t}}{t}.
\end{equation}
Assuming $\lambda,t>0$, a surface-localized solution (for $y<0$)
requires $\mathrm{Re}(\xi_{\zeta,\pm})>0$, which selects $\zeta=-1$.The
boundary solutions are then given by:
\begin{align}
|\psi_{1}\rangle & =\frac{1}{\sqrt{\mathcal{N}}}(e^{\xi_{-,+}y}-e^{\xi_{-,-}y})\phi_{-}^{1},\nonumber \\
|\psi_{2}\rangle & =\frac{1}{\sqrt{\mathcal{N}}}(e^{\xi_{-,+}y}-e^{\xi_{-,-}y})\phi_{-}^{2},\label{eq:zero_energy_solution}
\end{align}
with the normalization factor as $\mathcal{N}=\frac{(\xi_{-,+}-\xi_{-,-})^{2}}{2\xi_{-,+}\xi_{-,-}(\xi_{-,+}+\xi_{-,-})}$.
Here, the spinor components are defined as $\phi_{-}^{1}=|\rho_{y}=1\rangle\otimes|\sigma_{y}=-1\rangle$
and $\phi_{-}^{2}=|\rho_{y}=-1\rangle\otimes|\sigma_{y}=1\rangle$.
The matrix elements between these states are:
\begin{equation}
\langle\psi_{2}|\rho_{x}\sigma_{x}|\psi_{1}\rangle=1,\langle\psi_{2}|\rho_{x}\sigma_{z}|\psi_{1}\rangle=i.
\end{equation}
The altermagnetic term acts as a perturbation. Its matrix element
is:
\begin{equation}
\langle\psi_{2}|-\partial_{y}^{2}\rho_{z}\sigma_{z}|\psi_{1}\rangle=\xi_{-,+}\xi_{-,-}=\frac{2(m-3t)}{t},
\end{equation}
indicating that the altermagnetic term shifts the momentum of the
edge states by
\begin{equation}
k_{AM}=-\frac{t_{AM}}{\lambda}\frac{(m-3t)}{t}.
\end{equation}
Finally, projecting the remaining terms $\mathcal{H}_{1}$ onto the
basis $\{|\psi_{1}\rangle,|\psi_{2}\rangle\}$ yields the effective
surface Hamiltonian:
\begin{equation}
\mathcal{H}_{surf}=\lambda(-k_{x}+k_{AM})\sigma_{x}-\lambda k_{z}\sigma_{y}.
\end{equation}
In the presence of superconductivity, the corresponding BdG Hamiltonian
is given by:
\begin{equation}
\mathcal{H}_{\mathrm{BdG}}=\lambda(-k_{x}\tau_{z}+k_{AM}\tau_{0})\sigma_{x}-\lambda k_{z}\tau_{z}\sigma_{y}+\Delta\tau_{x},
\end{equation}
where $\tau_{i}$ are Pauli matrices acting in particle-hole space.
For $k_{z}=0$, the eigenenergies are 
\begin{equation}
E_{s,\zeta}=s\lambda k_{AM}+\zeta\sqrt{\Delta^{2}+(\mu-s\lambda k_{x})^{2}}\label{eq:edge_state_with_SC-1}
\end{equation}
with $s,\zeta=\pm1$. This spectrum indicates that the surface states
become nodal when the altermagnetic strength exceeds a critical value,
$t_{AM}>t_{c}^{s}$, where 
\begin{equation}
t_{c}^{s}\equiv\frac{\Delta}{|m-3t|}t.\label{eq:surface_critical_strength}
\end{equation}
Figure \ref{fig:surface_nodal_transition} shows the band structure
$E(k_{x},k_{z})$ at $k_{z}=\pi$ for a system with open boundaries
along the $y$-direction. The black dashed lines represent the analytical
solutions from Eq. (\ref{eq:edge_state_with_SC-1}) (black dashed
line). The numerical results show excellent agreement with the analytical
model in both the subcritical $t>t_{c}^{s}$ and supercritical $t<t_{c}^{s}$
regimes.

Unlike bulk BFSs which are protected by $\mathcal{I}\Xi$ symmetry,
surface BFSs arise from the topological surface states of the parent
AMTI proximitized to $s$-wave superconductivity. The normal-state
AMTI belongs to class DIII, characterized by a $\mathbb{Z}$ topological
invariant, i.e., the 3D winding number $w_{3D}$. On specific crystal
facets where the altermagnetic order acts as a momentum-dependent
Zeeman field (e.g., (100) and (010) surfaces), the Dirac cones are
shifted in an anisotropic way. When superconductivity is added, the
combination of spin splitting and topological protection prevents
full gapping, leading to gapless surface BFSs. Their existence is
therefore protected by the bulk topology of the underlying AMTIs.

Both bulk and surface BFSs are robust against weak disorder that preserves
(or statistically restores) the relevant protecting symmetries. For
bulk BFSs, disorder that respects $\mathcal{I}\Xi$ symmetry on average
does not gap out the nodal surfaces. For surface BFSs, disorder that
does not close the bulk gap of the parent AMTI and does not break
the crystalline symmetries essential for facet-dependent anisotropy
(e.g., $C_{4z}\mathcal{T}$) leaves the surface states intact. From
the perspective of the self-consistent Born approximation, weak scalar
(non-magnetic) disorder primarily renormalizes band parameters such
as effective mass and chemical potential, without qualitatively altering
the topological character. This renormalization does not destroy the
BFSs as long as the disorder strength remains below the threshold
that drives a topological phase transition.

\section{Energy scales and experimental feasibility}

In candidate altermagnetic materials such as $\mathrm{Eu}\mathrm{In}_{2}\mathrm{As}_{2}$,
the momentum-dependent spin splitting can reach values on the order
of 150\,meV \citep{cuono2023ab}. In our effective tight-binding model,
$t_{AM}$ corresponds to half the maximum of the spin-splitting energy
between opposite spin bands along high-symmetry directions, i.e. $t_{AM}\sim75\,$meV.

The surface critical scale is given by Eq. (\ref{eq:surface_critical_strength})
as $t_{c}^{s}=\frac{\Delta}{m_{bulk}}t$ where $2m_{bulk}=2|m-3t|$
is the bulk band gap protecting the topological surface states. The
parameter $t$ characterizes the spin independent hopping between
neighboring sites and is typically on the order of $\mathcal{O}(1\,\mathrm{eV})$.
In $\mathrm{Eu}\mathrm{In}_{2}\mathrm{As}_{2}$, $2m_{bulk}\sim0.1\,$eV
\citep{XuY19prl}. The pairing amplitude $\Delta$ is typically on
the order of $0.1-1$ meV, which is substantially smaller than all
other electronic energy scales in the system ($\Delta\ll m,t$). Assuming
$\Delta\approx0.2\,$meV and $t\approx1\,$eV yields $t_{c}^{s}\approx2\,$meV.
The bulk critical scale from Eq.\,(\ref{eq:bulk_critical_strength})
simplifies when the chemical potential lies within the bulk gap ($\mu\sim0$):
$t_{c}^{b}\sim\frac{1}{2}|m-t|\sim1\,$eV. Thus, the altermagnetic
strength $t_{AM}$ naturally satisfies the hierarchy of the energy
scales

\begin{equation}
t_{c}^{s}<t_{AM}<t_{c}^{b},
\end{equation}
that is fundamental to our key proposals of this work.

The inequality stems from the distinct physical origins of the two
critical scales: $t_{c}^{s}$ is pairing-limited ($t_{c}^{s}\propto\Delta\cdot(t/m_{bulk})$),
while $t_{c}^{b}$ is bandwidth-limited ($t_{c}^{b}\sim|m-t|$). The
pairing strength $\Delta$ is an order of magnitude smaller than the
bandwidth. Moreover, the intermediate altermagnetic energy $t_{AM}$
falls between these two extremes. Consequently, the unconventional
topological phenomena predicted in this work\textemdash both in nanowires
and in vortex lines\textemdash are readily accessible without fine-tuning
in heterostructures combining AMTIs and conventional superconductors.

\section{Topological phase transition driven by altermagnetism in the quasi-1D
nanowire}

\begin{figure}
\includegraphics[width=10cm]{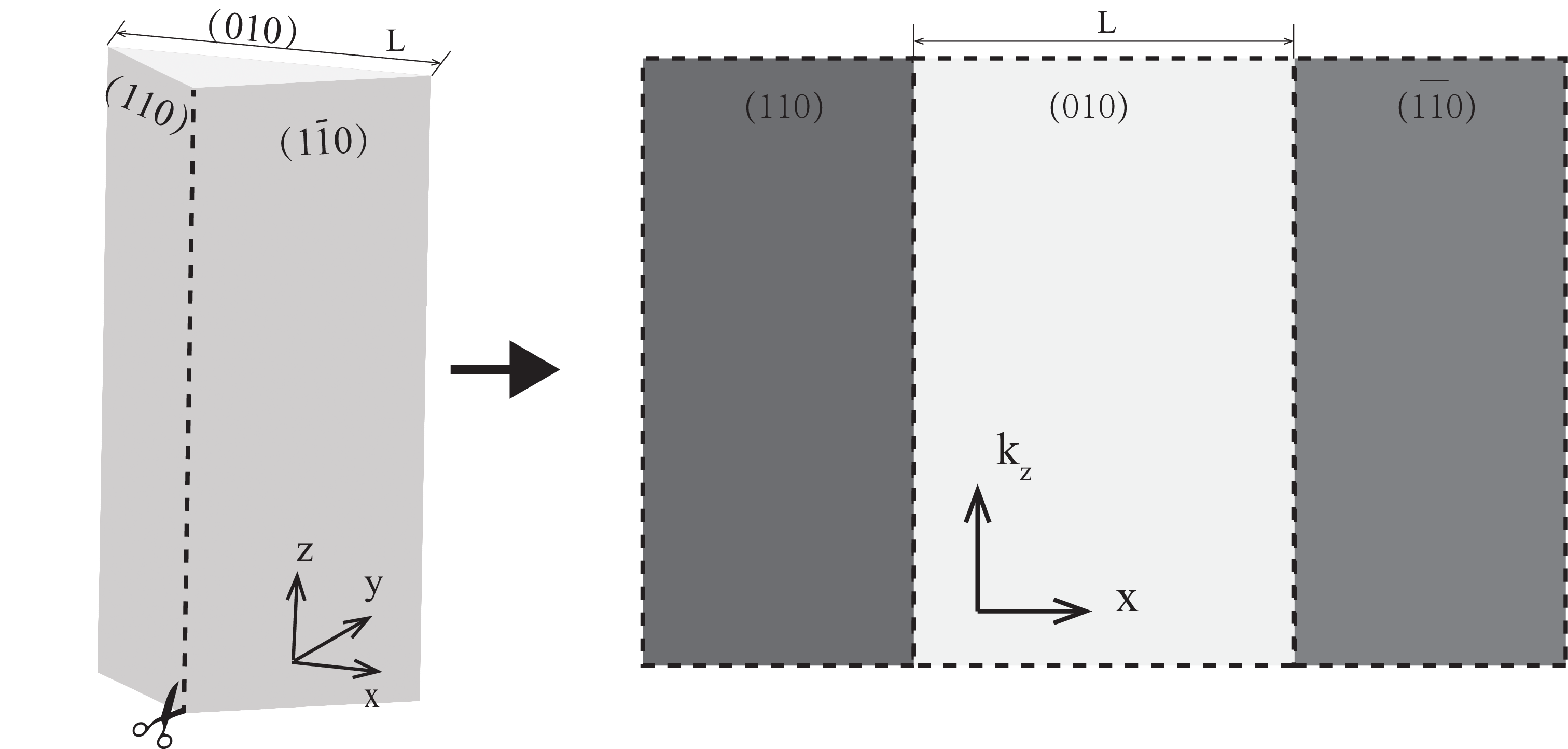}\caption{Diagram illustrating the unfolding of the prism's surface. \label{fig:schematic_diagram}}
\end{figure}

To investigate the topological properties of the system, we analyze
the quasiparticle spectrum within a triangular prism geometry, shown
schematically in Fig. \ref{fig:schematic_diagram}. The prism features
three relevant facets: (110) and (1$\bar{1}$0) and (010). The long
edge of the prism is aligned parallel to the $x$-axis (of length
$L$), allowing $k_{z}$ to remain a good quantum number. By conceptually
cutting along the interface between the (110) and (1$\bar{1}$0) surfaces
and unfolding them onto a plane, we effectively map the problem to
a strip geometry. This strip consists of a central region formed by
the (010) surface, sandwiched between the two other surfaces. The
key mechanism is the surface-projected altermagnetism: it induces
a Zeeman-like splitting in the (010) surface states but vanishes on
the (110) and (1$\bar{1}$0) surfaces. This allows us to write an
effective low-energy Hamiltonian for the strip:
\begin{align}
\mathcal{H} & =\mu\tau_{z}-i\lambda\partial_{x}\tau_{z}\sigma_{x}+\lambda k_{z}\tau_{z}\sigma_{y}+\lambda k_{AM}(x)\tau_{0}\sigma_{x}+\Delta\tau_{x},
\end{align}
where the altermagnetic momentum $k_{AM}(x)$ is spatially modulated:
\begin{equation}
k_{AM}(x)=\begin{cases}
k_{AM}, & -\frac{L}{2}<x<\frac{L}{2};\\
0, & \mathrm{elsewhere}.
\end{cases}
\end{equation}
In this framework, we solve an eigenvalue problem to obtain the spectrum
of discrete energy levels. For energies within the superconducting
gap $|\epsilon|<\Delta$, there are no propagating states in the gapped
(110) and (1$\bar{1}$0) regions outside the central strip. Consequently,
particles incident on the interfaces at $x\LyXThinSpace=\LyXThinSpace\pm\LyXThinSpace L/2$
undergo only normal and Andreev reflection without transmissions.
Since the bulk gap of the continuum model closes at $k_{z}=0$ as
the altermagnetic strength increases, we set $k_{z}=0$ to identify
the condition for gap inversion. This defines the critical boundaries
between topological and trivial phases in the $t_{AM}-L$ parameter
space.

For $k_{z}=0$, the Hamiltonian commutes with $\sigma_{x}$, allowing
us to decouple the problem into two independent sectors labeled by
the eigenvalue $s=\pm1$.

We first find the solutions in the different regions:

(a) Central $(010)$ surface region ($|x|<L/2$, $k_{AM}\ne0$):

Within this region, for a given sector $s$, the Hamiltonian reduces
to:
\begin{equation}
\mathcal{H}_{s}=\mu\tau_{z}-is\lambda\partial_{x}\tau_{z}+\lambda sk_{AM}\tau_{0}+\Delta\tau_{x}.
\end{equation}
We seek solutions of the form $\psi=\phi e^{i\kappa x}$, where $\phi$
is a two-component spinor in particle-hole space. Solving the eigenvalue
equation $\mathcal{H}_{s}\psi_{s}=\varepsilon\psi_{s}$ yields the
complex wavevectors:
\begin{equation}
\lambda\kappa_{s,\zeta}^{M}=\zeta\sqrt{(\varepsilon-s\lambda k_{AM})^{2}-\Delta^{2}}-s\mu,\,\,\,\zeta=\pm1.
\end{equation}
The corresponding eigenfunctions are: 
\begin{equation}
|\psi_{s,\zeta}^{M}\rangle=\left(\begin{array}{c}
\frac{|\varepsilon-s\lambda k_{AM}|+s\zeta\sqrt{(\varepsilon-s\lambda k_{AM})^{2}-\Delta^{2}}}{\Delta}\\
1
\end{array}\right)e^{i\kappa_{s,\zeta}^{M}x}.
\end{equation}
(b) Left $(110)$ surface region ($x<-L/2,k_{AM}=0$):

Here, the projection of altermagnetic term vanishes. We require bound
state solutions that decay exponentially as $x\to-\infty$. The physically
admissible wavevector is:
\begin{align}
\lambda\kappa_{s}^{L} & =-i\sqrt{\Delta^{2}-\varepsilon^{2}}-s\mu,
\end{align}
The corresponding eigenfunction is:
\begin{equation}
|\psi_{s}^{L}\rangle=\left(\begin{array}{c}
\frac{-si\sqrt{\Delta^{2}-\varepsilon^{2}}+\varepsilon}{\Delta}\\
1
\end{array}\right)e^{(\sqrt{\Delta^{2}-\varepsilon^{2}}-si\mu)x/\lambda}.
\end{equation}
The imaginary part of $\kappa_{s}^{L}$ ensures exponential decay.

(c) Right (1$\bar{1}$0) surface region ($x>L/2$ and $k_{AM}=0$):

Similarly, for decay as $x\to\infty$, the wavevector is:
\begin{equation}
\lambda\kappa_{s}^{R}=i\sqrt{\Delta^{2}-\varepsilon^{2}}-s\mu.
\end{equation}
The corresponding eigenfunction is:
\begin{equation}
|\psi_{s}^{R}\rangle=\left(\begin{array}{c}
\frac{si\sqrt{\Delta^{2}-\varepsilon^{2}}+\varepsilon}{\Delta}\\
1
\end{array}\right)e^{(\sqrt{\Delta^{2}-\varepsilon^{2}}-si\mu)x/\lambda}.
\end{equation}
We then construct the full wavefunction and apply boundary conditions.
The general wavefunction in the central region is a superposition
of the left- and right-propagating (or evanescent) solutions:
\begin{equation}
|\Psi_{s}^{M}\rangle=\sum_{\zeta=\pm}A_{\zeta}|\psi_{s,\zeta}^{M}\rangle.
\end{equation}
We now focus on the $s=+1$ channel. The bound state spectrum is found
by demanding that the wavefunctions are continuous at the two interfaces,
$x=\pm L/2$. This leads to a system of equations:
\begin{align}
|\Psi_{s}^{M}(-\frac{L}{2})\rangle & =B|\psi_{s}^{L}(-\frac{L}{2})\rangle,\\
|\Psi_{s}^{M}(\frac{L}{2})\rangle & =C|\psi_{s}^{R}(\frac{L}{2})\rangle.\nonumber 
\end{align}
Here, $B$ and $C$ are the amplitudes of the decaying waves in the
left and right regions, respectively.

Next, we find the secular equation and analysis topological phase
transition. This system of equations can be written in the matrix
form $\mathbf{M}\mathbf{v}=0$ where $\mathbf{v}=(A_{+},A_{-},B,C)^{T}$.
For a non-trivial solution to exist, the determinant of the coefficient
matrix $\mathbf{M}$ must be zero: $det(\mathbf{M})=0$. Solving this
determinant condition yields the dispersion relation for discrete
energy levels:
\begin{equation}
\tan\left(L\sqrt{(\lambda k_{AM}-\varepsilon)^{2}-\Delta^{2}}\right)=\frac{\sqrt{(\Delta^{2}-\varepsilon^{2})((\lambda k_{AM}-\varepsilon)^{2}-\Delta^{2})}}{\varepsilon|\lambda k_{AM}-\varepsilon|-\Delta^{2}}.
\end{equation}
The gap inversion and the emergence of topological states are signaled
by the appearance of zero-energy modes $(\varepsilon=0)$. Substituting
$\varepsilon=0$ into the dispersion relation yields a simplified
equation: $\tan\left((\Delta L)\sqrt{(\frac{\lambda k_{AM}}{\Delta})^{2}-1}\right)=-\sqrt{(\frac{\lambda k_{AM}}{\Delta})^{2}-1}$
which admits solutions for the system size
\begin{equation}
L/\xi=\frac{n\pi-\arctan\sqrt{(k_{AM}\xi)^{2}-1}}{\sqrt{(k_{AM}\xi)^{2}-1}},
\end{equation}
where $\xi=\lambda/\Delta$ is the superconducting coherence length
and $n=1,2,3,\cdots$. This equation defines the phase boundary in
the ($t_{AM}$, $L$) parameter space. Solutions to this equation
exist only when $\lambda k_{AM}>\Delta$, confirming that the altermagnetic
strength must exceed the superconducting gap to drive the system into
the topological phase. The size $L$ of the (010) facet then determines
the precise quantization of the bound state energy and the number
of topological modes.
\begin{figure}
\includegraphics[width=16cm]{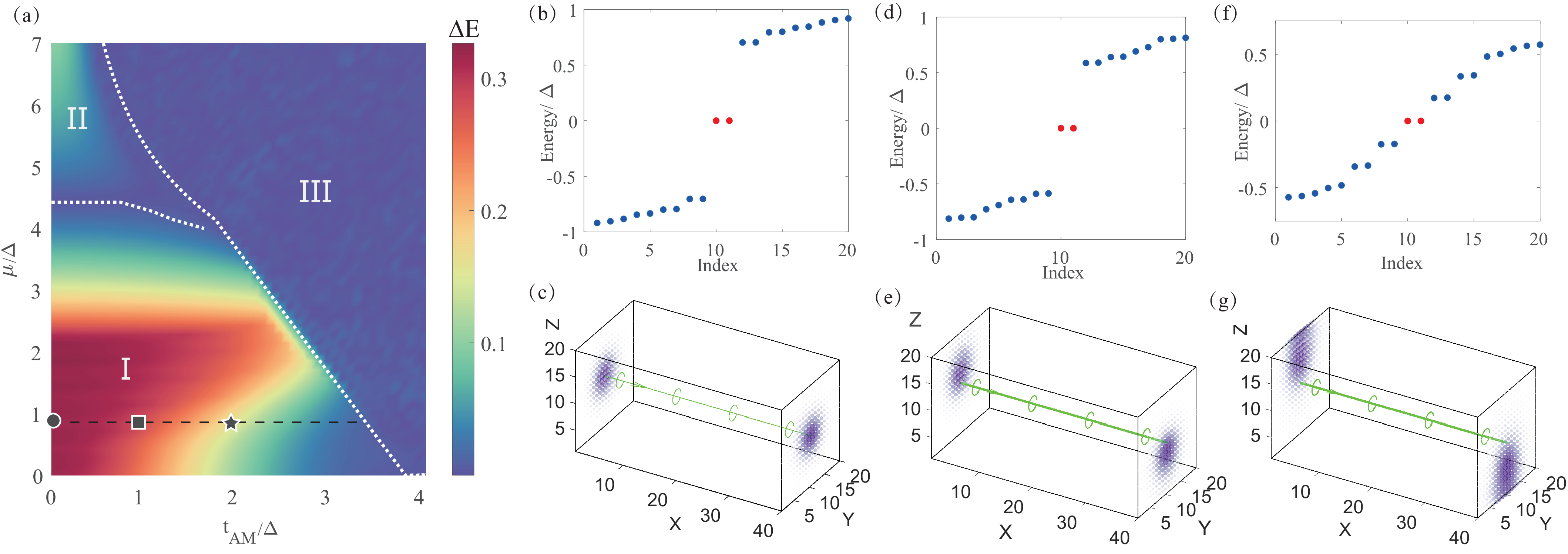}

\caption{(a)Topological phase diagram for a vortex line along the $x$-axis
in the ($t_{AM}/\Delta,\mu/\Delta$) parameter space. The phases are
labeled as follows: I. Topological vortex line phase, II. Trivial
vortex phase, and III. Nodal phase. The color scale represents the
minigap $\Delta E$. The black dashed line at $\mu/\Delta=1$ indicates
the parameter cut for panels (b, d, f), with the sphere, square, and
hexagon marking the specific values of $t_{AM}/\Delta=0,1,2$, respectively.
The white dashed line indicates the critical boundary. (b, d, f) Energy
spectra under open boundary conditions in the x-direction for $t_{AM}/\Delta=0,1,2,$
respectively. Majorana zero modes are highlighted by red dots. (c,
e, g) Spatial profiles of the zero-energy wavefunction magnitude for
$t_{AM}/\Delta=0,1,2,$ respectively. The evolution shows the Majorana
state shifting from the vortex core towards the sample hinge as the
altermagnetic strength increases. Parameters: $\Delta=0.2$, $m=2.5$.\label{fig:vortex_along_x}}
\end{figure}

\section{Spatial shift of Majorana zero mode in a vortex line along the x-direction}

In this section, we investigate a vortex line oriented along the $x$-direction,
where $k_{x}$ remains a good quantum number. Figure \ref{fig:vortex_along_x}(a)
maps the minigap of this quasi-1D system in the ($t_{AM}/\Delta,\mu/\Delta$)
parameter space. The phase diagram reveals three distinct regions:
I. topological vortex line phase, II. trivial vortex phase, and III.
nodal phase. Notably, unlike the vortex along the $z$-direction discussed
in the main text, the $x$-direction vortex lacks the oscillating
gap regime (previously labeled IV) . To understand the evolution of
the topological phase, we trace a path at a fixed chemical potential
of $\mu/\Delta=1$, indicated by the black dashed line in \ref{fig:vortex_along_x}(a).
As the altermagnetic strength $t_{AM}$ increases, the energy gap
gradually closes, culminating in a nodal phase for $t_{AM}>t_{c}^{b}$
(.the white dashed line marking the phase boundary to the nodal phase
(III)). Crucially, within the gapped phase, the system is characterized
by a $\mathbb{Z}_{2}$ topological invariant, protected by parity
and particle-hole symmetry, which ensures the persistence of MZMs.

We now examine how the altermagnetic order modifies the properties
of MZMs. Figures \ref{fig:vortex_along_x}(b, d, f) show the energy
spectra with open boundary conditions along $x$-direction for increasing
altermagnetic strength $t_{AM}$ ($t_{AM}/\Delta=0$, $1$, $2$),
corresponding to the sphere, square, and hexagon markers in (a). The
MZMs are highlighted by red dots. Although the bulk gap diminishes
with increasing $t_{AM}$, the MZM remains robust at zero energy as
long as the gap is finite.

The key effect of the AM field is revealed in the spatial profile
of the MZM wavefunctions, shown in Figs. \ref{fig:vortex_along_x}(c,
e, g). At $t_{AM}=0$, the MZM is tightly localized at the vortex
core. As $t_{AM}$ increases, the center of the MZM wavefunction shifts
away from the core. At a sufficiently strong $t_{AM}$, the MZM relocates
to the sample hinge. This spatial shift can be understood intuitively:
The altermagnetic order acts as a momentum-dependent Zeeman field.
On surfaces parallel to the $x$-direction, this combines with the
physical magnetic field of the vortex (also along $x$ direction).
The point where the net effective field vanishes\textemdash the natural
binding site for the MZM\textemdash is consequently displaced away
from the geometric center. Due to the alternating sign of the altermagnetic
field, adjacent side surfaces experience opposite effective Zeeman
fields, resulting in opposite shifts of the MZMs on the front and
back terminations.

The deviation of the MZMs from the vortex core provides a direct experimental
signature for detecting AM order in superconducting topological systems.
This distinct spatial shift serves as a definitive marker to distinguish
altermagnetic vortices from conventional superconducting ones and
offers a novel mechanism for manipulating MZMs in topological quantum
devices.

\section{$g$-wave Altermagnetic topological insulator in a candidate material
$\mathrm{Eu}\mathrm{In_{2}As_{2}}$}

\subsection{Crystal symmetry and magnetic structure}

\begin{figure}
\includegraphics[width=12cm]{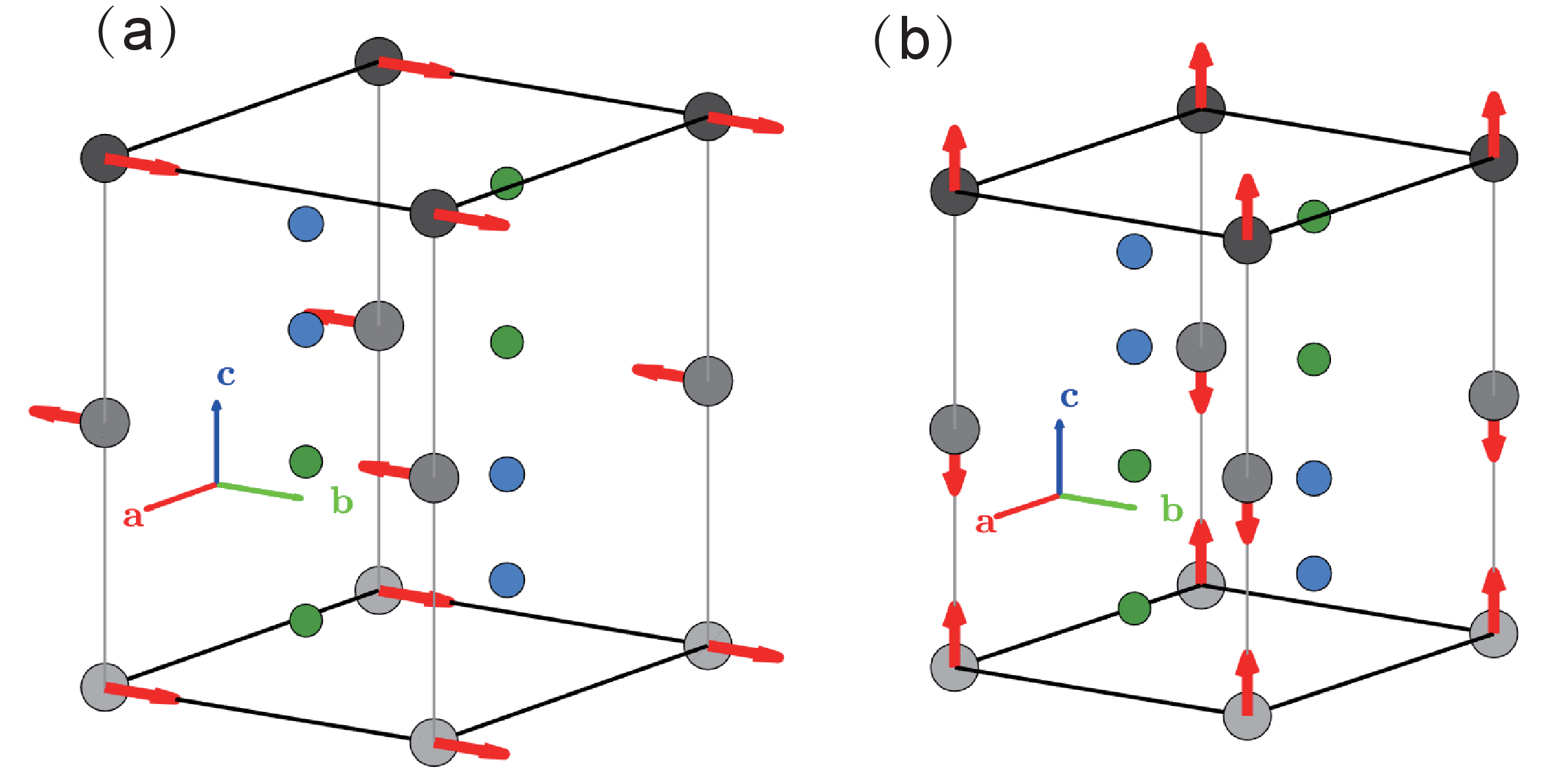}\caption{Crystal and magnetic structure of EuIn$_{2}$As$_{2}$. (a) afmb phase
with Néel vector along the $b$-axis. (b) afmc phase with Néel vector
along the $c$-axis. The crystal axes $b$ and $c$ are oriented along
the $y$ and $z$ directions of the Cartesian coordinate system, respectively.
Black, blue, and green circles represent Eu, In, and As atoms, respectively.
Red arrows indicate the magnetic moments on Eu atoms, showing the
antiferromagnetic ordering along the Néel vector direction.\label{fig:Crystal-and-magnetic}}
\end{figure}

The candidate material $\mathrm{Eu}\mathrm{In_{2}As_{2}}$ crystallizes
in the hexagonal space group $P6_{3}/mmc$ (No. 194), consisting of
alternating hexagonal layers of europium (Eu) and $\mathrm{In_{2}As_{2}}$
blocks along the crystallographic $c-$axis as shown in Fig. \ref{fig:Crystal-and-magnetic}.
In this structure, Eu atoms occupy the $2a$ Wyckoff site at $(0,0,0)$
and $(0,0,1/2)$, while both In and As reside on the $4f$ Wyckoff
site with general coordinates: $\pm(2/3,1/3,z)$ and $\pm(1/3,2/3,z+1/2)$
in the basis of the unit cell axes $\mathbf{a}$, $\mathbf{b}$ and
$\mathbf{c}$. For convenience, we align the crystal axes with the
Cartesian coordinate system such that $\mathbf{b}||\hat{y}$ and $\mathbf{c}||\hat{z}$.
The space group $P6_{3}/mmc$ is generated by inversion ($I$), a
$2\pi/3$ rotation around $\hat{z}$ ($C_{3z})$, a $\pi$ rotation
around $\hat{z}$ followed by a $c/2$ translation along $\hat{z}$
($\{C_{2z}|00\frac{c}{2}\}$) and a $\pi$ rotation around {[}010{]}:
$C_{2y}$; note that the combination $C_{2y}\{C_{2z}|00\frac{c}{2}\}$
yields $\{C_{2x}|00\frac{c}{2}\}$, a $\pi$ rotation around $\hat{x}$
with the same half-translation.

In the absence of spin-orbit coupling, spin orientation is not irrelevant;
the collinear antiparallel spin order of $\mathrm{Eu}\mathrm{In}_{2}\mathrm{As}_{2}$
corresponds to the nontrivial spin Laue group $^{2}6/^{2}m^{2}m^{1}m([E||\bar{3}m]+[C_{2}||C_{6z}][E||\bar{3}m]$)
\citep{Libor22prx2}. However, when considering topological properties,
spin-orbit coupling becomes essential, and we must instead consider
the magnetic space group operations. In the paramagnetic phase, the
magnetic space group is $P6_{3}/mmc1^{\prime}$ (No. 194.264), a grey
group in which time-reversal symmetry $T$ accompanies every symmetry
operation of the crystallographic space group.

The material hosts two metastable antiferromagnetic (AFM) phases:
the afmb phase with Néel vector along the crystallographic $b$-axis
{[}Fig. \ref{fig:Crystal-and-magnetic}(a){]} and the afmc phase with
the Néel vector along the $c$-axis {[}Fig. \ref{fig:Crystal-and-magnetic}
(b){]} \citep{XuY19prl}. First-principles calculations show that
the two phases have similar energies, with a difference of less than
1 meV \citep{XuY19prl}. Table S1 summarizes the symmetry operations
preserved ($\text{\CheckmarkBold}$) or broken ($\text{\ding{56}}$)
in each phase. The distinct symmetry-breaking patterns underlie the
different topological properties of the two phases.
\begin{table}
\begin{tabular}{|c|c|c|c|c|c|c|c|c|c|c|c|}
\hline 
 & $C_{3z}$ & $\{C_{2z}|00\frac{c}{2}\}$ & $C_{2y}$ & $\{C_{2x}|00\frac{c}{2}\}$ & $I$ & $T$ & $TC_{3z}$ & $T\{C_{2z}|00\frac{c}{2}\}$ & $TC_{2y}$ & $T\{C_{2x}|00\frac{c}{2}\}$ & $TI$\tabularnewline
\hline 
\hline 
afmb & \ding{56} & \CheckmarkBold{} & \CheckmarkBold{} & \CheckmarkBold{} & \CheckmarkBold{} & \ding{56} & \CheckmarkBold{} & \ding{56} & \ding{56} & \ding{56} & \ding{56}\tabularnewline
\hline 
afmc & \CheckmarkBold{} & \ding{56} & \ding{56} & \CheckmarkBold{} & \CheckmarkBold{} & \ding{56} & \ding{56} & \CheckmarkBold{} & \CheckmarkBold{} & \ding{56} & \ding{56}\tabularnewline
\hline 
\end{tabular}\caption{Symmetry analysis of the two antiferromagnetic phases. The table shows
which symmetries are preserved (\CheckmarkBold ) or broken (\ding{56})
in the two metastable AFM configurations with Néel vectors aligned
along the crystallographic $b$-axis (afmb phase) and $c$-axis (afmc
phase). The symmetry operations include three-fold rotation $C_{3z}$,
two-fold rotations $\{C_{2z}|00\frac{c}{2}\}$, $C_{2y}$, and $\{C_{2x}|00\frac{c}{2}\}$
with associated translations, spatial inversion $I$, time-reversal
$T$, and their various combinations.}
\end{table}

In the afmb phase, the antiferromagnetic order breaks $C_{3z}$ symmetry
but preserves two mirror symmetries: $M_{y}=IC_{2y}$ and $M_{z}=I\{C_{2z}|00\frac{c}{2}\}$.
Consequently, we can define mirror Chern numbers at the time-reversal
invariant planes $k_{z}=0,\pi$ and $k_{y}=0,\pi$ , denoted $n_{M_{z}}$
and $n_{M_{y}}$, respectively. Nonzero mirror Chern numbers $n_{Mz}=\pm1$
($n_{My}=\pm1$) guarantee the existence of massless Dirac cones on
surfaces that preserve the corresponding mirror symmetry.

In the afmc phase, where the Néel vector aligns along the crystallographic
$c$-axis, the magnetic order preserves a unique combination of symmetries
that fundamentally constrain the low-energy electronic structure.
The crucial symmetry is the composition of three-fold rotation $C_{3z}$
with the time-reversal-screw operation $T\{C_{2z}|00\frac{c}{2}\}$,
which together generate an effective six-fold rotational symmetry
$T\{C_{6z}|00\frac{c}{2}\}$. We find that the $g$-wave altermagnetic
order is permitted by the combined symmetry operation $T\{C_{6z}|00\frac{c}{2}\}$
\citep{Libor22prx2}, which exists exclusively in the afmc phase.
The opposite spin sublattices are connected by the symmetries $M_{z}T$
and $M_{y}T$, with $M_{z}=I\{C_{2z}|00\frac{c}{2}\}$ and $M_{y}=IC_{2y}$.

\begin{table}
\begin{tabular}{|c|c|c|}
\hline 
Symmetry operation & Matrix representation & Momentum transformation\tabularnewline
\hline 
\hline 
$C_{3z}$ & $e^{-i\frac{\pi}{3}\sigma_{z}}$ & $(-\frac{1}{2}k_{x}+\frac{\sqrt{3}}{2}k_{y},-\frac{\sqrt{3}}{2}k_{x}-\frac{1}{2}k_{y},k_{z})$\tabularnewline
\hline 
$\{C_{2x}|00\frac{c}{2}\}$ & $e^{-i\frac{\pi}{2}\sigma_{x}}$ & $(k_{x},-k_{y},-k_{z})$\tabularnewline
\hline 
$I$ & $\rho_{z}$ & $-\mathbf{k}$\tabularnewline
\hline 
$T\{C_{2z}|00\frac{c}{2}\}$ & $\sigma_{x}K$ & $(k_{x},k_{y},-k_{z})$\tabularnewline
\hline 
$TC_{2y}$ & $K$ & $(k_{x},-k_{y},k_{z})$\tabularnewline
\hline 
\end{tabular}

\caption{Symmetry operations and their representations in the four-dimensional
basis for the afmc phase. For each operation, we list the corresponding
$4\times4$ matrix representation acting on the orbital ($\rho_{i}$)
and spin ($\sigma_{i}$ ) degrees of freedom, as well as the transformation
of the momentum vector $\mathbf{k}=(k_{x},k_{y},k_{z})$. The operator
$K$ denotes complex conjugation.}
\end{table}

\subsection{Low-energy effective Hamiltonian for afmc phase}

At the $\Gamma$ point, the two bands near the Fermi level transform
under the irreducible representations $\Gamma_{4}^{+}$ and $\Gamma_{4}^{-}$
of the double group $D_{3d}$. According to the character table of
$D_{3d}$, $\Gamma_{4}^{\pm}$ are two-dimensional representations
corresponding to states with total angular momentum $j_{z}=\pm3/2$
and definite parity under inversion. The complete set of symmetry
operations and their representations in this four-dimensional basis
are listed in Table S2. These symmetry operations impose stringent
constraints on the form of the Hamiltonian. Based on symmetry-invariant
theory, we construct the low-energy effective Hamiltonian around $\Gamma$
point expanded to fourth order in $k$ that respects all symmetries
of the afmc phase. The Hamiltonian takes the form:
\begin{align}
H(\mathbf{k}) & =C(\mathbf{k})I_{4}+\underbrace{M(\mathbf{k})\rho_{z}\sigma_{0}+vk_{x}\rho_{x}\sigma_{x}+vk_{y}\rho_{x}\sigma_{y}+v_{z}k_{z}\rho_{x}\sigma_{z}}_{BHZ}\nonumber \\
 & +D\rho_{z}[\sigma_{x}(k_{x}^{2}-k_{y}^{2})-2\sigma_{y}k_{x}k_{y}]+\underbrace{Jk_{z}(k_{x}^{3}-3k_{x}k_{y}^{2})\rho_{z}\sigma_{z}}_{g-wave-AM},\label{eq:bulk_H}
\end{align}
with the mass and energy dispersion functions:
\begin{align}
M(\mathbf{k}) & =m_{0}-m_{1}(k_{x}^{2}+k_{y}^{2})-m_{2}k_{z}^{2},\\
C(\mathbf{k}) & =C_{0}-C_{1}(k_{x}^{2}+k_{y}^{2})-C_{2}k_{z}^{2}.
\end{align}
The parameters $v$, $v_{z}$, $D$, and $J$ are real constants determined
by material-specific details. The physical interpretation of each
term is as follows: The Bernevig\textendash Hughes\textendash Zhang
(BHZ) terms describe a three-dimensional topological insulator with
spin-orbit coupling. The $C(\mathbf{k})$ term accounts for electron-hole
asymmetry, which does not affect the topology and is therefore neglected
in the following analysis. The $D$ term represents a nematic distortion
that breaks the in-plane rotational symmetry. The final term, proportional
to $J$, is the $g$-wave altermagnetic coupling that constitutes
the central focus of this work. This term respects all symmetries
of the afmc phase while generating a momentum-dependent spin splitting
with $g$-wave symmetry, characterized by the $k_{z}(k_{x}^{3}-3k_{x}k_{y}^{2})$
functional form.

\begin{figure}
\includegraphics[width=16cm]{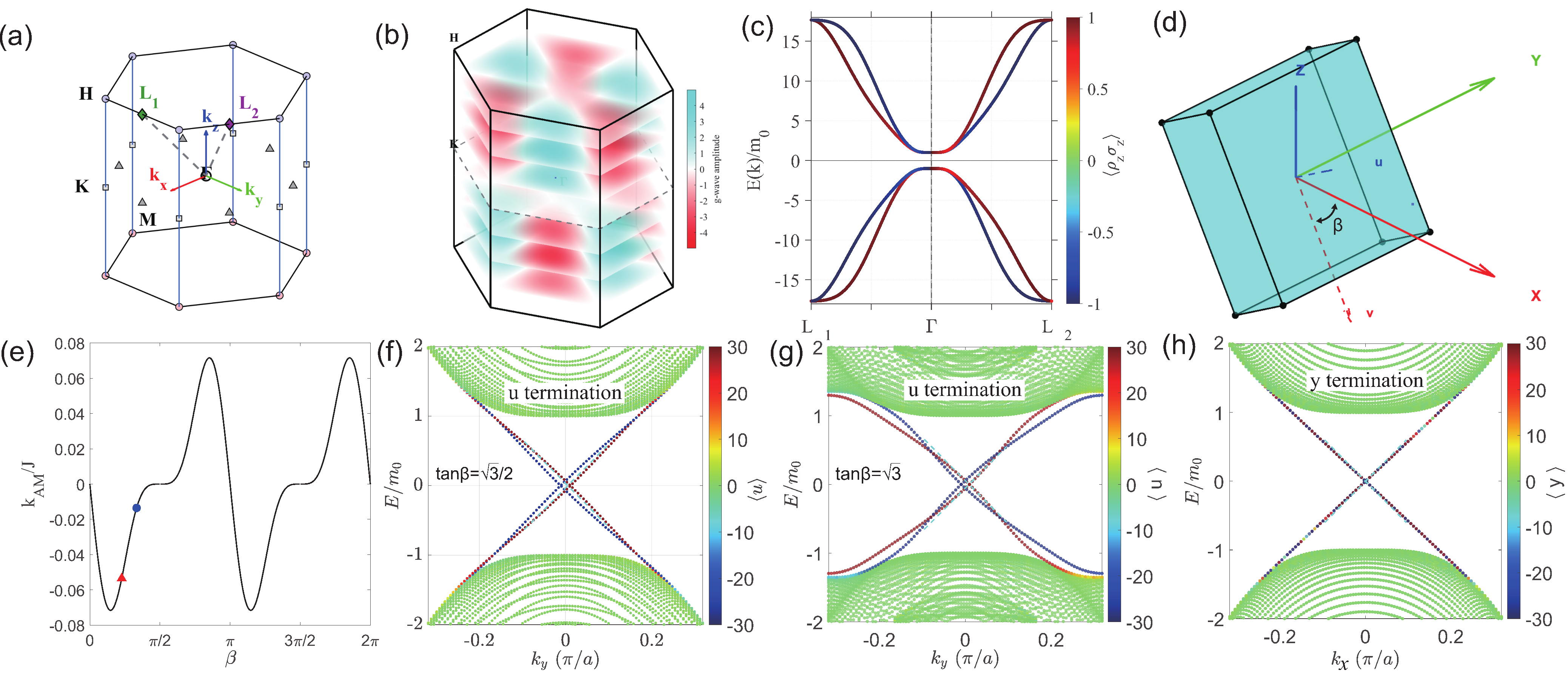}\caption{(a) First Brillouin zone of a typical hexagonal primitive cell with
the high-symmetry points highlighted in different symbols. The dashed
gray line indicates the high-symmetry momentum path used to show the
altermagnetism in space group $P6_{3}/mmc$. (b) Schematic illustration
of the $g$-wave altermagnetism in the first Brillouin zone, with
color scale representing the amplitude of the altermagnetic splitting
(red to blue indicates negative to positive values). (c) Band structure
of the bulk Hamiltonian in Eq. (\ref{eq:bulk_H}) along the high-symmetry
line $\mathrm{L}_{1}-\Gamma-\mathrm{L}_{2}$. The color scale from
blue to red indicates the spin polarization $\langle\rho_{z}\sigma_{z}\rangle$
for $J=0.3$. (d) Schematic illustration of the termination in the
$xz$ plane along the $u$ direction, which is the $x$ axis rotated
by angle $\beta$ about the $y$-axis; $u$ and $v$ are the normal
and parallel directions in this termination. (e) The Dirac surface
cone shift $k_{AM}$ for $u$ termination along the $k_{y}$ direction
induced by $g$-wave altermagnetism as a function of the termination
angle $\beta$. (f) Slab geometry band structure as a function of
$k_{y}$ with $k_{v}=0$ for $u$ termination with $\tan\beta=\sqrt{3}/2$
and $J=0.3$. (g) Same as (f) but with $\tan\beta=\sqrt{3}$ and $J=2$.
(h) $y$-termination slab band structure as a function of $k_{x}$
with $k_{z}=0$. Cyan dashed lines indicate the analytic solutions.
Other parameters: $m_{1}=m_{2}=v=v_{z}=1$, $m_{0}=0.5$, $D=0$.\label{fig:S5}}
\end{figure}

\subsection{Lattice regularization}

To enable numerical simulations, we regularize the continuum $\mathbf{k\cdot p}$
Hamiltonian on a three-dimensional hexagonal lattice. The direct lattice
vectors are defined as: $\mathbf{a}=\frac{1}{2}(\sqrt{3},-1,0)a$,
$\mathbf{b}=(0,1,0)a$, $\mathbf{c}=(0,0,1)c$, where $a$ and $c$
are the in-plane and out-of-plane lattice constants. The corresponding
reciprocal lattice vectors are: $\mathbf{b}_{1}=2\pi(\frac{2}{\sqrt{3}a},0,0)$,
$\mathbf{b}_{2}=2\pi(-\frac{1}{\sqrt{3}a},\frac{1}{a},0)$, and $\mathbf{b}_{3}=2\pi(0,0,\frac{1}{c})$.
The Brillouin zone (BZ) forms a hexagonal prism {[}Fig. \ref{fig:S5}(a){]}
with high-symmetry points $\mathbf{L}_{1}=2\pi(\frac{1}{\sqrt{3}a},0,\frac{1}{2c})$
and $\mathbf{L}_{2}=2\pi(-\frac{1}{2\sqrt{3}a},-\frac{1}{2a},\frac{1}{2c})$.
To construct a lattice model that faithfully reproduces the low-energy
continuum Hamiltonian, we replace continuum momentum operators with
their lattice analogues through the following substitutions:

Linear terms: 
\begin{align}
\mathbf{k}\cdot\boldsymbol{\sigma} & \to\frac{2}{3a}\sum_{i=1,2,3}\sin(\mathbf{k}\cdot\mathbf{a}_{i})(\hat{\mathbf{a}}_{i}\cdot\boldsymbol{\sigma})+\frac{1}{c}\sin(\mathbf{k}\cdot\mathbf{c})(\hat{\mathbf{c}}\cdot\boldsymbol{\sigma}).
\end{align}
where $\hat{\mathbf{a}}_{i}=\mathbf{a}_{i}/|\mathbf{a}_{i}|$ and
$\hat{\mathbf{c}}=\mathbf{c}/|\mathbf{c}|$ are unit vectors.

Quadratic terms:
\begin{align}
k_{x}^{2}+k_{y}^{2} & \to\frac{4}{3a^{2}}[3-\sum_{i=1,2,3}\cos(\mathbf{k}\cdot\mathbf{a}_{i})], & k_{z}^{2}\to\frac{2}{c^{2}}[1-\cos(\mathbf{k}\cdot\mathbf{c})].
\end{align}

$d$-wave terms:
\begin{align}
\sigma_{x}(k_{x}^{2}-k_{y}^{2})-\sigma_{y}2k_{x}k_{y} & \to-\frac{8}{3a^{2}}\sum_{i}\cos(\mathbf{k}\cdot\mathbf{a}_{i})(\hat{\mathbf{a}}_{i}\times\boldsymbol{\sigma})_{z}.
\end{align}

Quartic altermagnetic term: The $g$-wave term requires regularization
beyond nearest neighbors. We introduce a set of next-nearest hopping
vectors in $ab$ plane: $\mathbf{a}_{1}^{\prime}=(1,0,0)\sqrt{3}a,\mathbf{a}_{2}^{\prime}=\frac{1}{2}(-1,-\sqrt{3},0)\sqrt{3}a,$
and $\mathbf{a}_{3}^{\prime}=-\mathbf{a}_{1}^{\prime}-\mathbf{a}_{2}^{\prime}$.
Using these vectors, the cubic term regularizes as:
\begin{equation}
k_{x}^{3}-3k_{x}k_{y}^{2}\to\frac{8}{3\sqrt{3}a^{3}}\sum_{i}\sin(\mathbf{k}\cdot\mathbf{a}_{i}^{\prime}).
\end{equation}
The $g$-wave altermagnetic spin splitting is shown schematically
in Fig. \ref{fig:S5}(b), which changes sign under a $C_{6z}$ transformation
and under $k_{z}\to-k_{z}$.

Using the regularized operators above, we obtain the lattice Hamiltonian
in momentum space:
\begin{align}
H & =C_{l}(\mathbf{k})I_{4}+M_{l}(\mathbf{k})\rho_{z}\sigma_{0}\nonumber \\
 & +\rho_{x}\left[\frac{2v}{3a}\sum_{i=1,2,3}\sin(\mathbf{k}\cdot\mathbf{a}_{i})(\hat{\mathbf{a}}_{i}\cdot\boldsymbol{\sigma})+\frac{v_{z}}{c}\sin(\mathbf{k}\cdot\mathbf{c})(\hat{\mathbf{c}}\cdot\boldsymbol{\sigma})\right]\nonumber \\
 & +\frac{J}{c}\sin(k_{z}c)\frac{8}{3\sqrt{3}a^{3}}\sum_{i=1,2,3}\sin(\mathbf{k}\cdot\mathbf{a}_{i}^{\prime})\rho_{z}\sigma_{z},
\end{align}
where the lattice-regularized mass and energy dispersion functions
are:
\begin{align}
M_{l}(\mathbf{k}) & =\widetilde{m}_{0}+m_{1}\frac{4}{3a^{2}}\sum_{i=1,2,3}\cos(\mathbf{k}\cdot\mathbf{a}_{i})+m_{2}\frac{2}{c^{2}}\cos(\mathbf{k}\cdot\mathbf{c}),\nonumber \\
C_{l}(\mathbf{k}) & =\widetilde{C}_{0}+C_{1}\frac{4}{3a^{2}}\sum_{i=1,2,3}\cos(\mathbf{k}\cdot\mathbf{a}_{i})+C_{2}\frac{2}{c^{2}}\cos(\mathbf{k}\cdot\mathbf{c}),
\end{align}
with renormalized parameters:
\begin{align}
\widetilde{m}_{0} & =m_{0}-m_{1}\frac{4}{a^{2}}-\frac{2}{c^{2}}m_{2}, & \widetilde{C}_{0}=C_{0}-C_{1}\frac{4}{a^{2}}-\frac{2}{c^{2}}C_{2}.
\end{align}
We plot the eigenspectrum along the high-symmetry line $\mathrm{L}_{1}-\Gamma-\mathrm{L}_{2}$
{[}gray dashed line in Fig. \ref{fig:S5}(a){]}. As shown in Fig.
\ref{fig:S5}(c), the spin splitting reverses sign along $\mathrm{L}_{1}-\Gamma$
and $\mathrm{L}_{2}-\Gamma$, revealing the characteristic $g$-wave
altermagneic nature \citep{cuono2023ab,pari2025nonrelativistic}.
For real-space simulations, we Fourier transform the momentum-space
Hamiltonian to obtain a tight-binding model:
\begin{align}
H & =\sum_{\mathbf{r}}\{\widetilde{m}_{0}\psi_{\mathbf{r}}^{\dagger}\rho_{z}\sigma_{0}\psi_{\mathbf{r}}+\sum_{i=1,2,3}\psi_{\mathbf{r}+\mathbf{a}_{i}}^{\dagger}(m_{1}\frac{2}{3a^{2}}\rho_{z}\sigma_{0}+\frac{v}{3ai}\rho_{x}\hat{\mathbf{a}}_{i}\cdot\boldsymbol{\sigma})\psi_{\mathbf{r}}\nonumber \\
 & +\psi_{\mathbf{r}+\mathbf{c}}^{\dagger}(\frac{m_{2}}{c^{2}}\rho_{z}\sigma_{0}+\frac{v_{c}}{2ic}\rho_{x}\sigma_{z})\psi_{\mathbf{r}}+\frac{J}{c}\frac{2}{3\sqrt{3}a^{3}}\sum_{i=1,2,3}(\psi_{\mathbf{r}+\mathbf{a}_{i}^{\prime}+\mathbf{c}}^{\dagger}\rho_{z}\sigma_{z}\psi_{\mathbf{r},z}-\psi_{\mathbf{r}+\mathbf{a}_{i}^{\prime}-\mathbf{c}}^{\dagger}\rho_{z}\sigma_{z}\psi_{\mathbf{r},z})+h.c.\}.
\end{align}

\subsection{Surface states on a specific termination}

We now analyze the low-energy surface states that arise when the $g$-wave
AMTI is terminated along a specific crystallographic direction. The
total Hamiltonian in the bulk is $H=H_{BHZ}+H_{AM}^{g}$. We consider
a surface termination perpendicular to the $xz$-plane as shown in
Fig. \ref{fig:S5}(d). To handle the open boundary condition, we perform
a rotation in the $xz$-plane to define a new coordinate system ($u,v,y$)
where $u$ is the direction normal to the surface, and $v$ is an
in-plane coordinate. The rotation is defined by an angle $\beta$
around the $y$-axis: $u=\cos\beta x+\sin\beta z$ and $v=-\sin\beta x+\cos\beta z$.
The corresponding momentum transformation is:
\begin{equation}
k_{x}=\cos\beta k_{u}+\sin\beta k_{v},k_{z}=-\sin\beta k_{u}+\cos\beta k_{v}.
\end{equation}
Substituting these transformations, the altermagnetic term becomes:
\begin{align}
H_{AM}^{g} & =J(-\sin\beta k_{u}+\cos\beta k_{v})(\cos\beta k_{u}+\sin\beta k_{v})((\cos\beta k_{u}+\sin\beta k_{v})^{2}-3k_{y}^{2})\rho_{z}\sigma_{z}.
\end{align}
For a surface termination along the $u$-direction, translational
invariance is preserved in the $y$ and $v$ directions, allowing
$k_{y},k_{v}$ to be good quantum numbers. In the low-energy, long-wavelength
limit relevant to bound states, we expand to leading order in $k_{y}$
and $k_{v}$, (i.e., $k_{y},k_{v}\approx0$). The altermagnetic term
then simplifies to a $k_{u}^{4}$ term
\begin{align}
H_{AM}^{g} & \simeq-J\sin\beta\cos^{3}\beta k_{u}^{4}\rho_{z}\sigma_{z}.\label{eq:AM_along_u}
\end{align}
 Next, we transform the BHZ Hamiltonian. After substituting the momentum
transformation, we obtain:
\begin{align}
H_{BHZ} & =k_{u}(v\cos\beta\rho_{x}\sigma_{x}-v_{z}\sin\beta\rho_{x}\sigma_{z})+k_{v}(v\sin\beta\rho_{x}\sigma_{x}+v_{z}\cos\beta\rho_{x}\sigma_{z})+vk_{y}\rho_{x}\sigma_{y}\nonumber \\
 & +(m_{0}-m_{u}k_{u}^{2}-m_{v}k_{v}^{2}-2m_{uv}k_{u}k_{v}-m_{1}k_{y}^{2})\rho_{z}\sigma_{0}
\end{align}
where the effective masses in the rotated frame are
\begin{align}
m_{u} & =m_{1}\cos^{2}\beta+m_{2}\sin^{2}\beta,\nonumber \\
m_{v} & =m_{1}\sin^{2}\beta+m_{2}\cos^{2}\beta,\nonumber \\
m_{uv} & =(m_{1}-m_{2})\cos\beta\sin\beta.
\end{align}
To simplify the linear $k$-terms, we perform a unitary rotation in
spin space, $U_{\theta}=e^{i\sigma_{y}\frac{\theta}{2}}$, to diagonalize
the velocity operator along the $u$-direction. The angle $\theta$
is defined as 
\begin{equation}
\cos\theta=\frac{v\cos\beta}{v_{u}},\sin\theta=\frac{v_{z}\sin\beta}{v_{u}}
\end{equation}
with $v_{u}=\sqrt{v^{2}\cos^{2}\beta+v_{z}^{2}\sin^{2}\beta}$. Applying
this rotation, the Hamiltonian becomes:

\begin{align}
U_{\theta}H_{BHZ}U_{\theta}^{-1} & =v_{u}k_{u}\rho_{x}\sigma_{x}+k_{v}\rho_{x}(\frac{v_{z}v}{v_{u}}\sigma_{z}+\frac{v^{2}-v_{z}^{2}}{2v_{u}}\sin2\beta\sigma_{x})+vk_{y}\rho_{x}\sigma_{y}\nonumber \\
 & +(m_{0}-m_{u}k_{u}^{2}-m_{v}k_{v}^{2}-2m_{uv}k_{u}k_{v}-m_{1}k_{y}^{2})\rho_{z}\sigma_{0}+\cos\theta\sigma_{z}-\sin\theta\sigma_{x}.
\end{align}
Under the same unitary transformation, the altermagnetic term in Eq.
(\ref{eq:AM_along_u}) transforms to:
\begin{equation}
U_{\theta}H_{AM}^{g}U_{\theta}^{-1}=-J\sin\beta\cos^{3}\beta k_{u}^{4}\rho_{z}(\frac{v\cos\beta}{v_{u}}\sigma_{z}-\frac{v_{z}\sin\beta}{v_{u}}\sigma_{x}).\label{eq:transformed_AM}
\end{equation}
We now focus on the surface bound states by solving the Hamiltonian
at $k_{v}=k_{y}=0$. The Hamiltonian simplifies to a 1D Dirac equation
in $u$:
\begin{equation}
H_{0}=-iv_{u}\partial_{u}\rho_{x}\sigma_{x}+(m_{0}+m_{u}\partial_{u}^{2})\rho_{z}\sigma_{0}.
\end{equation}
We seek evanescent solutions decaying into the bulk ($u<0$) of the
form $|\psi_{i}\rangle\propto(e^{\xi_{-,+}u}-e^{\xi_{-,-}u})|\phi_{-}^{i}\rangle$,
where $\xi_{-,\pm}=\frac{v_{u}\pm\sqrt{v_{u}^{2}-4m_{0}m_{u}}}{2m_{u}}$
and the spinor basis states are: $|\phi_{-}^{1}\rangle=|\rho_{y}=-1\rangle\otimes|\sigma_{x}=1\rangle$
and $|\phi_{-}^{2}\rangle=|\rho_{y}=1\rangle\otimes|\sigma_{x}=-1\rangle$.
Finally, we project the $k_{v}$ and $k_{y}$ dependent terms from
$H_{1}=H-H_{0}$ onto this surface state subspace. The effective 2D
Hamiltonian for the surface states is:
\begin{equation}
H_{surf}=-\frac{v_{z}v}{v_{u}}k_{v}\sigma_{y}-v(k_{y}-k_{AM})\sigma_{x},\label{eq:projection_surface_state_in_u}
\end{equation}
where the altermagnetic term generates a momentum-dependent shift
$k_{AM}$. This shift is calculated from the matrix element of the
transformed altermagnetic term {[}Eq. (\ref{eq:transformed_AM}){]}
between the two surface states:
\begin{equation}
vk_{AM}=\frac{v\cos\beta}{v_{u}}\langle\psi_{1}|-J\sin\beta\cos^{3}\beta\partial_{u}^{4}\rho_{z}\sigma_{z}|\psi_{2}\rangle
\end{equation}
Crucially, although the altermagnetic term vanishes at the Dirac point
($E=0$), the coupling to surface states involves a higher-order $\partial_{u}^{4}$
term. This quartic derivative, arising from the lattice regularization
of the $g$-wave altermagnetism, produces a finite shift $k_{\mathrm{AM}}$
even at zero energy, resulting in a tunable Dirac cone shift controlled
by the surface orientation $\beta$. Evaluating this matrix element
using the explicit wavefunctions, we find
\begin{equation}
vk_{AM}=J\frac{\tan\beta}{\sqrt{1+\frac{v_{z}^{2}}{v^{2}}\tan^{2}\beta}}\frac{m_{0}\left((m_{0}m_{1}-v^{2})+(m_{0}m_{2}-v_{z}^{2})\tan^{2}\beta\right)}{(m_{1}+m_{2}\tan^{2}\beta)^{3}}.\label{eq:kam_vs_beta}
\end{equation}
It shows that the altermagnetism induces a momentum-dependent spin-orbit
field that shifts the Dirac cone of the surface states, with a strength
tunable by the material parameters and the surface orientation angle
$\beta$.

As shown in \ref{fig:S5}(e), we plot $k_{AM}$ as a function of $\beta$.
When $\beta=0,\frac{\pi}{2},\pi,\frac{3}{2}\pi$ that the surface
normal aligns with the crystal axes, $k_{AM}=0$, indicating that
the altermagnetism projected onto these terminations vanishes. We
then plot the band structure of a slab geometry with periodic boundary
conditions along the $y$ and $v$ directions. As shown in Fig. \ref{fig:S5}(f,
g) for two $\beta$ values with $\tan\beta=\sqrt{3}/2$ {[}red triangle
in Fig. \ref{fig:S5}(e){]} and $\tan\beta=\sqrt{3}$ {[}black circle
in (e){]}, the Dirac surface states for termination perpendicular
to $xz$-plane are shifted along $k_{y}$. The cyan dashed lines represent
the analytical expression from Eq. (\ref{eq:projection_surface_state_in_u})
which shows good agreement with the numerical results. Note that in
Fig. \ref{fig:S5}(g), we use a larger $J$ value to clearly visualize
the shift. As shown in Fig. \ref{fig:S5}(h), we also show the surface
states for $y$ termination as a function of $k_{x}$. In this case,
the projection of altermagnetism vanishes, so no shift is observed.

\subsection{Topological Superconductivity in $g$-wave Altermagnetic Topological
Insulators}

\begin{figure}
\includegraphics[width=16cm]{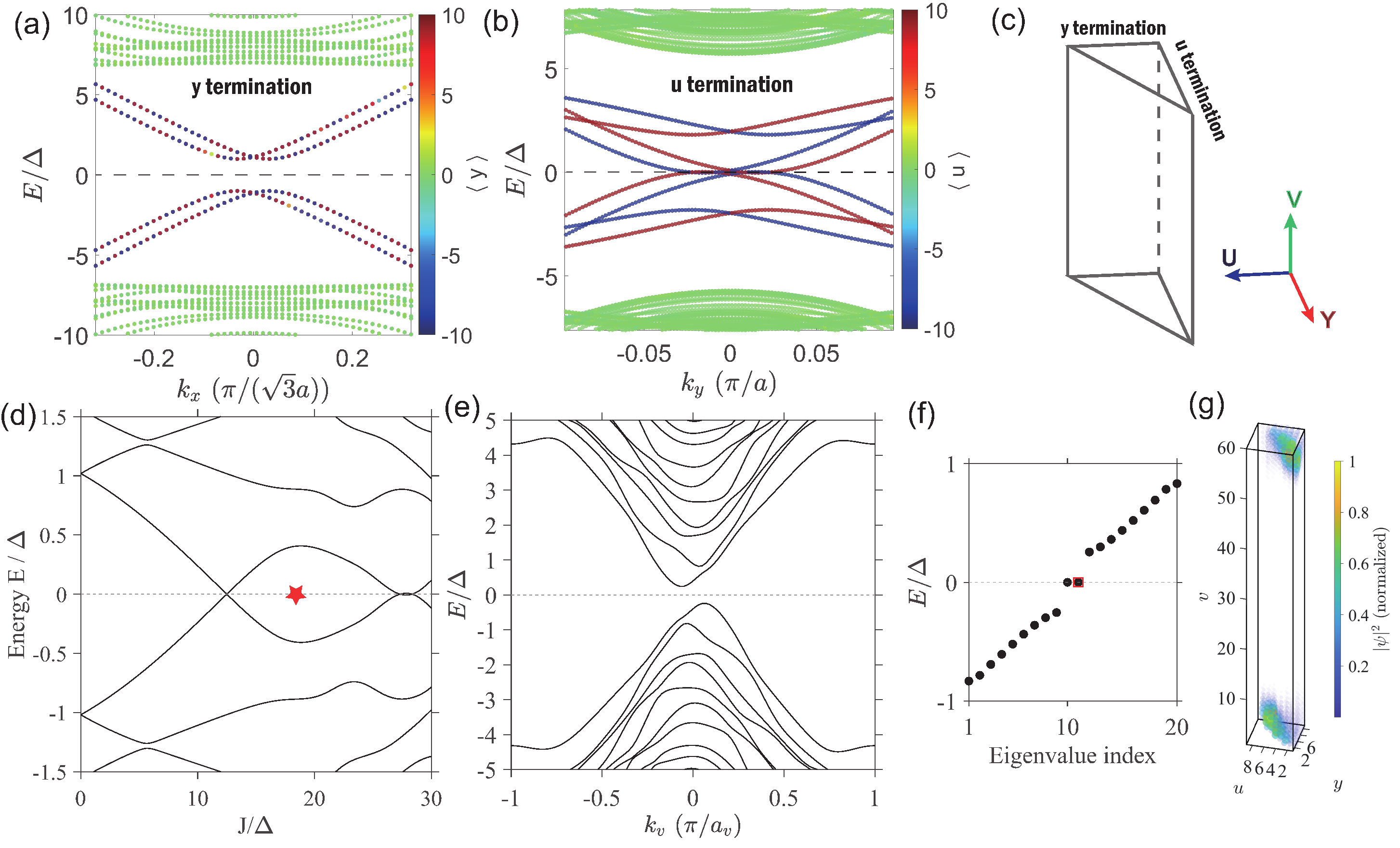}

\caption{(a) Slab band structure for $y$-termination with superconductivity
$\Delta$ as a function of $k_{x}$. (b) Slab band structure for $u$-termination
with $\tan\beta=\sqrt{3}/2$ as a function of $k_{y}$. (c) Schematic
diagram of the triangular prism used to realize topological superconductivity.
(d) Quasi-1D band structure with open boundaries in $u$ and $y$
directions and $k_{v}$ as a good quantum number, plotted as a function
of $J/\Delta$ at $k_{v}=0$. (e) Quasi-1D band structure as a function
of $k_{v}$ for $J/\Delta=18$ (denoted by the red pentagram in (d)).
(f) Spectrum under open boundary condition along $v$ for $J/\Delta=18$.
(g) Wave function distribution of the zero-energy solution (highlighted
by the red box in (f)). The color gradient from yellow to blue represents
the amplitude of the wavefunction. Parameters: $m_{1}=m_{2}=v=v_{z}=1$,
$m_{0}=0.8$, $D=0$\label{fig:gwAMTI_with_SC}}
\end{figure}

The interplay between $g$-wave altermagnetism and superconductivity
in a topological insulator provides a promising platform for realizing
topological superconductivity. As shown in Fig. \ref{fig:gwAMTI_with_SC}(a),
the $y$-termination slab exhibits a conventional superconducting
gap. In contrast, the $u$-termination slab with $\tan\beta=\sqrt{3}/2$
{[}Fig. \ref{fig:gwAMTI_with_SC}(b){]} reveals a distinctive feature:
due to the finite momentum shift induced by $g$-wave altermagnetism,
the Dirac surface states fail to develop a fully gapped superconducting
spectrum, resulting instead in a Bogoliubov Fermi surface.

To explore the emergence of topological superconductivity, we consider
a triangular prism geometry {[}Fig.\ref{fig:gwAMTI_with_SC}(c){]}
that combines open boundaries in the $uy$-plane while preserving
translational invariance along $v$. This geometry effectively creates
a quasi-one-dimensional system where the altermagnetic field can induce
nontrivial topological phases. Figure \ref{fig:gwAMTI_with_SC}(d)
shows the quasi-1D band structure as a function of the altermagnetic
coupling strength $J/\Delta$ at $k_{v}=0$. As $J$ increases, a
clear crossing of states occurs, a hallmark of a topological phase
transition. The nature of the spectrum at a point after the phase
transition {[}red pentagram in Fig. \ref{fig:gwAMTI_with_SC}(d){]}
is further elucidated in Fig. \ref{fig:gwAMTI_with_SC}(e), which
reveals a fully gapped bulk spectrum. The full spectrum under open
boundary conditions along $v$ {[}Fig. \ref{fig:gwAMTI_with_SC}(f){]}
confirms the presence of a robust zero-energy state, marked by the
red box, well isolated from the bulk continuum. The spatial profile
of this zero-energy state is displayed in Fig. \ref{fig:gwAMTI_with_SC}(g).
The wavefunction exhibits characteristic exponential decay away from
the boundaries, with amplitude concentrated at the top and bottom
surfaces of the triangular geometry. This localization pattern, combined
with the topological protection of the zero-energy mode, strongly
suggests the formation of Majorana bound states.

Thus, our results demonstrate that the $g$-wave altermagnetic topological
insulator provides a tunable platform for engineering Majorana bound
states, with the surface orientation $\beta$ and altermagnetic coupling
$J$ serving as key control parameters. In summary, we have demonstrated
that the interplay between $g$-wave altermagnetism, superconductivity,
and a topological insulator substrate in the hexagonal space group
$P6_{3}/mmc$ provides a promising platform for realizing topological
superconductivity.

\end{document}